\documentclass[fleqn,10pt,twocolumn]{wlscirep}
\usepackage[utf8]{inputenc}
\usepackage[T1]{fontenc}

\DeclareUnicodeCharacter{2010}{-}   % hyphen
\DeclareUnicodeCharacter{2011}{-}   % non-breaking hyphen
\DeclareUnicodeCharacter{2012}{--}  % figure dash
\DeclareUnicodeCharacter{2013}{--}  % en dash
\DeclareUnicodeCharacter{2014}{---} % em dash
\DeclareUnicodeCharacter{2018}{`}   % left single quote
\DeclareUnicodeCharacter{2019}{'}   % right single quote
\DeclareUnicodeCharacter{00A0}{ }   % non-breaking space

\usepackage{bm}
\usepackage{multicol}
\usepackage{float}
\usepackage[skins,breakable]{tcolorbox}
\usepackage{stfloats}  % in preamble
\floatstyle{boxed}
\newfloat{mybox}{htbp}{lob}
\floatname{mybox}{Box}
\usepackage{soul}

\usepackage{tcolorbox}

\newtcolorbox{cvbox}[1][]{%
    enhanced, 
    breakable=true, % Make the box breakable (i.e., allows content to span multiple pages)
    after skip=8mm, % Adds vertical space after the box
    fonttitle=\sffamily\bfseries, % Title font style
    coltitle=cyan, % Title color
    colbacktitle=gray!10, % Background color for the title
    titlerule=0pt, % No rule underneath the title (optional, can be customized)
    colback=gray!5, % Box background color (light gray for readability)
    colframe=black!75, % Box frame color (dark gray)
    title=#1, % The title of the box is the first argument passed to the cvbox environment
    overlay={%
        \ifcase\tcbsegmentstate
        \or
        \else
        \fi%
    }
}

\usepackage{amsfonts}
\usepackage{soul}

\title{
Self-Organising Memristive Networks as Physical Learning Systems}

\author[1,2*]{Francesco Caravelli}
\author[3]{Gianluca Milano}
\author[4]{Adam Stieg}
\author[5]{Carlo Ricciardi}
\author[6]{Simon A. Brown}
\author[7]{Zdenka Kuncic}
\affil[1]{Theoretical Division (Condensed Matter and Complex Systems, T-4), Los Alamos National Laboratory, Los Alamos, New Mexico 87545, USA}
\affil[2]{Dipartimento di Fisica dell'Universit\`a di Pisa, Largo Bruno Pontecorvo 3, I-56127 Pisa, Italy}
\affil[3]{Advanced Materials Metrology and Life Sciences Division, INRiM (Istituto Nazionale di Ricerca Metrologica), Strada delle Cacce 91, 10135 Torino, Italy}
\affil[4]{University of California, Los Angeles, California NanoSystems Institute, Los Angeles, CA 90095, USA}
\affil[5]{Department of Applied Science and Technology, Politecnico di Torino, C.so Duca degli Abruzzi 24, 10129 Torino, Italy}
\affil[6]{The MacDiarmid Institute for Advanced Materials and Nanotechnology, School of Physical and Chemical Sciences, Te Kura Mate, University of Canterbury, Private Bag 4800, Christchurch 8140, New Zealand}
\affil[7]{School of Physics, Sydney Nano Institute and Centre for Complex Systems, University of Sydney, Sydney, NSW 2006, Australia}

\affil[*]{e-mail: caravelli@lanl.gov}

\begin{document}

\begin{abstract}
Learning with physical systems is an emerging paradigm that seeks to harness the intrinsic nonlinear dynamics of physical substrates for learning. The impetus for a paradigm shift in how hardware is used for computational intelligence stems largely from the unsustainability of artificial neural network software implemented on conventional transistor-based hardware.
This Perspective highlights one promising approach using physical networks comprised of resistive memory nanoscale components with dynamically reconfigurable, self-organising electrical circuitry. Experimental advances have revealed the non-trivial interactions within these Self-Organising Memristive Networks (SOMNs), offering insights into their collective nonlinear and adaptive dynamics, and how these properties can be harnessed for learning using different hardware implementations. Theoretical approaches, including mean-field theory, graph theory, and concepts from disordered systems, reveal deeper insights into the dynamics of SOMNs, especially during transitions between different conductance states where criticality and other dynamical phase transitions emerge in both experiments and models. Furthermore, parallels between adaptive dynamics in SOMNs and plasticity in biological neuronal networks suggest the potential for realising energy-efficient, brain-like continual learning.
SOMNs thus offer a promising route toward embedded edge intelligence, unlocking real-time decision-making for autonomous systems, dynamic sensing, and personalised healthcare, by embedding continuous learning in resource-constrained environments.
The overarching aim of this Perspective is to show how the convergence of nanotechnology, statistical physics, complex systems, and self-organising principles offers a unique opportunity to advance a new generation of physical intelligence technologies.
\end{abstract}

\thispagestyle{empty}

\flushbottom
\maketitle

\section{Introduction}

Physical learning systems can be broadly described as systems comprised of physical substrates whose intrinsic nonlinear dynamics can be exploited for learning \cite{Kaspar2021,Jaeger_2023}. Examples include analog electrical circuits, photonic and opto--electronic systems, as well as stochastic spintronic systems \cite{SternMurugan2022}.
A particularly promising approach aims to harness native brain--like, or neuromorphic, properties exhibited by certain nanoscale electronic devices, such as {memristors} (see \textbf{Box 1}), with physical memory and neuromorphic dynamics \cite{Wang_2020}. 
{Among the most promising examples of such systems are Self-Organising Memristive Networks (SOMNs). These physical networks are described as self--organising because under external electrical stimulation, they can spontaneously acquire specific novel structural and functional properties~\cite{Haken1988}, effectively making them {dynamically reconfigurable} electrical circuits. These macroscale, observable properties emerge from the collective dynamical interactions between their constituent nanoscale elements, in the spirit of ``More is Different''~\cite{Anderson1972}. As highlighted in this Perspective, SOMNs are unique {physical} learning substrates because their ability to encode, store, and process information arises directly from their material and structural properties, positioning them at the intersection of statistical physics, neuromorphic computing, nanotechnology, and complex systems.}

The importance of the structure and connectivity of biological neuronal networks {for} learning was {already} recognised by some of the earliest work on artificial intelligence~\cite{Caianiello1961,rosenblatt1958perceptron,rosenblatt1962principles}.
Rosenblatt's perceptron introduced probabilistic learning and statistical separability as central to cognition; he envisioned the perceptron as a biologically--inspired model for emergent dynamics from random but structured \emph{physical} wiring, while also noting practical limitations in scaling such architectures~\cite{rosenblatt1958perceptron,rosenblatt1962principles,hopfield1982neural,cybenko1989approximation}.
{These early insights motivated the first experimental realisations of SOMNs as scalable physical networks with brain--like architectures and dynamics achieved using disordered nano-electronic materials~\cite{Stieg2011,Avizienis2013,Stieg2014_self-org,Sandouk2015,Mallinson2019}.}

\begin{figure*}[ht]
    \centering
  \includegraphics[width=0.77\linewidth]{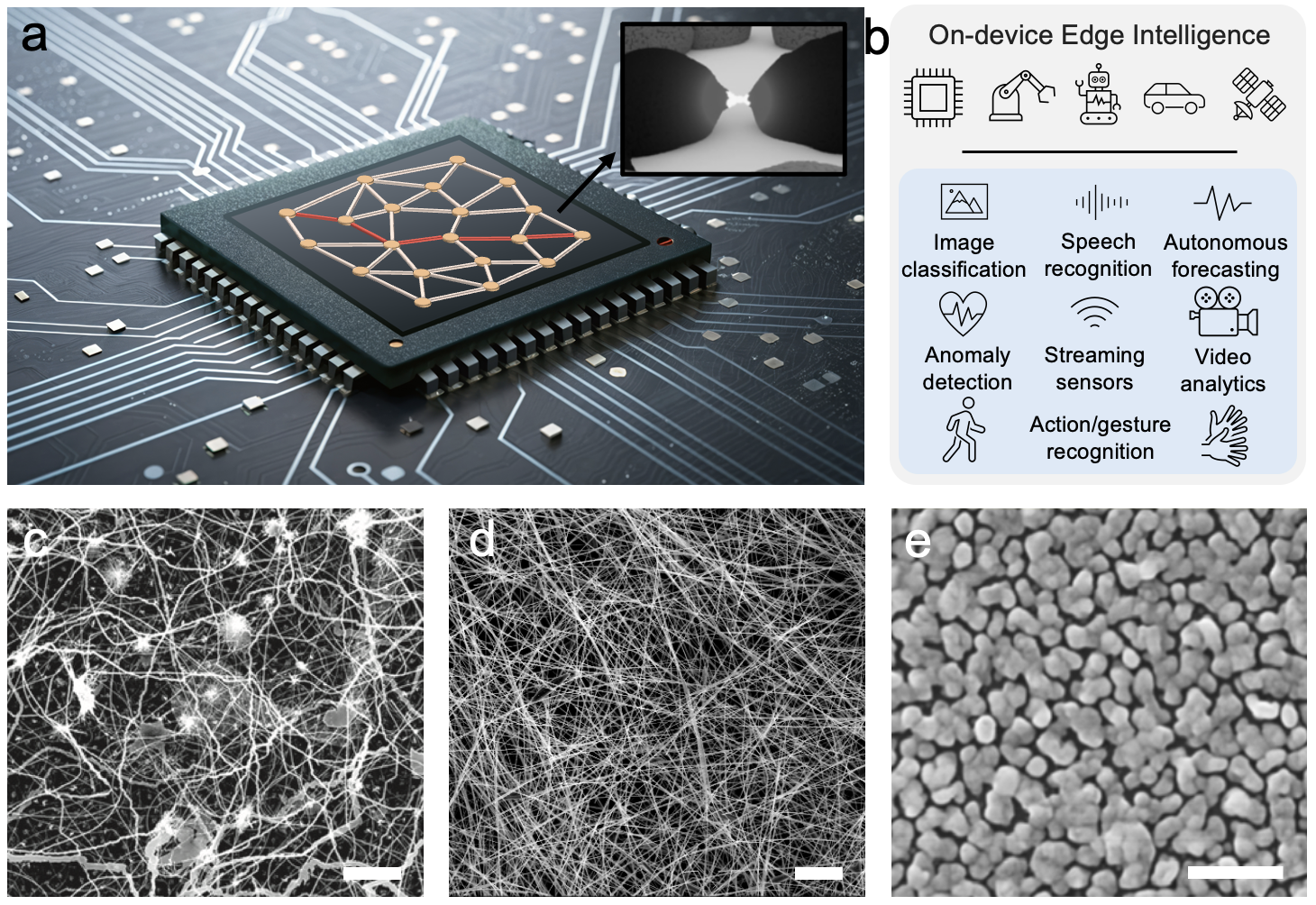}

 \caption{
Overview of self-organising memristive networks (SOMNs) as a platform for physical learning systems.
\textbf{a.} Conceptual figure of an on--chip SOMN device illustrating that, in response to electrical signals, the physical network substrate undergoes adaptive changes in conductance (red);
this is due to resistive switching in nanoscale junctions between {the self-assembled elements}, associated with atomic rearrangements and electron transport (inset) driven by the applied electric field \cite{Stieg2011,Milano2019,Vahl2024}
\textbf{b.} Envisaged applications and use cases at the sensor edge, where latency, bandwidth and compute resource constraints demand \textit{in situ} machine intelligence that is low power and energy efficient, and able to learn continuously on-device as the physical environment changes.
\textbf{c -- e.} Scanning Electron Microscope (SEM) images of different nanowire (c and d) and nanoparticle (e) SOMNs.
Scalebars are 25\,$\mu$m, 5\,$\mu$m and 200\,nm for panels c, d and e, respectively. SEM images in panels c and e are adapted from \cite{demis2015atomic} and \cite{mallinson2019avalanches}, respectively, {while d is original}.
}
    \label{fig:evo}
\end{figure*}

{In neuromorphic and brain--inspired computing, a current focus of research is the exploration of materials and devices that can be engineered to mimic the dynamics and efficiency of biological {neurons and synapses~\cite{Christensen_2022, Wang_BIC_2026}. In addition, memristor devices connected in a cross-bar array architecture map directly onto the bipartite graph structure of artificial neural networks (ANNs) and can thus serve as a physical instantiation of matrix--vector multiplications for efficient inference of ANN models \cite{xia2019memristive,Momeni2024}.}
In this Perspective we posit that SOMNs
are unique physical learning substrates because they mimic the structure--function relation of the brain, which is the archetypal non--equilibrium dynamical learning system \cite{Breakspear_brain-dynamics_2017,Lynn-Bassett_2019,DiazAlvarez2019,Loeffler2020topological,Suarez_2021,milano2022connectome,Vahl2024}. 
Typically comprised of nanowires~\cite{Stieg2011,Avizienis2013,demis2015atomic} (NWs) or nanoparticles~\cite{Kim_nanoparticle_2009,Sandouk2015,Mallinson2019} (NPs), SOMNs are formed via bottom--up synthesis, which naturally produces a high density of interconnections, far surpassing the limits of conventional top-down lithographic fabrication~\cite{Stieg2014_self-org}.
Their electronic architecture is also inherently stochastic~\cite{milano2025self}, just like biological neuronal networks, i.e. no two single brains are identical.

Figure~\ref{fig:evo} summarises the key {features} of SOMNs as {multiterminal} platforms for physical learning.
{SOMNs exhibit internal conductance states that adaptively respond to external electrical signals (Fig. \ref{fig:evo}~(a)), which may be delivered directly when embedded at the sensor edge for a wide range of applications (Fig. \ref{fig:evo}~(b)). Most importantly, the response is associated with internal dynamical states. Fig.~\ref{fig:evo}c,d and e show, respectively, scanning electron microscopy (SEM) images revealing the complex topology of SOMNs based on electrochemically grown NWs, randomly dispersed NWs, and randomly deposited NPs.}
In contrast to conventional hardware computing systems (and many unconventional ones), SOMNs can process information efficiently without the need for direct addressability of individual components \cite{Kaspar2021}, similarly to biological neural networks.
SOMNs can operate as physical learning substrates by applying relatively simple learning schemes that leverage their adaptive dynamical responses to input electrical signals, similar to biological plasticity and in contrast to global weight training used in ANNs. SOMNs can be viewed as complex physical systems, as they produce higher-order nonlinear dynamics,  criticality and phase transitions in the order parameter.
These non-trivial dynamics emerge from the interplay between nanoscale transport dynamics (which generate high degrees of freedom in their internal resistance states) and heterogeneous, recurrent network connectivity~\cite{DiazAlvarez2019,zhu2021information}.

This Perspective is intended for researchers across several disciplines. 
Physicists can view SOMNs as physical substrates to study emergent phenomena \cite{Anderson1972}, like criticality, by means of  dynamical systems  and statistical mechanics frameworks. Material scientists and chemists may be motivated to explore new types of functional nanomaterials with similar properties (e.g. dynamic reconfigurability, long-range spatio--temporal correlations, as well as resistance to perturbations and self--healing, all key properties of adaptive, complex systems \cite{Terasa}). Computational neuroscientists may find value in SOMNs as an abiotic platform for gaining new insights into how coupled, physics--constrained synaptic dynamics produce multi--scale brain-like dynamics and process information \cite{Chialvo1999,Carbajal2015,Carbajal2022,Barrows2024}. Computer scientists may instead view these {physical} systems as a potential platform for efficient machine learning applications and for exploring new approaches to solving computing problems. 

\begin{figure*}
\begin{tcolorbox}[title=Box 1: Memristive devices, colback=cyan!5!white, colframe=cyan!75!black, width=\textwidth]
The ideal memristor is defined by the relationship between flux ($\Phi$) and charge ($q$):
\begin{equation}
d\Phi = M \, dq,
\end{equation}
where $M$, the memristance, depends on $\Phi$ or $q$. If $M$ is constant, the device behaves as a linear resistor; if $M$ varies, the memristor exhibits nonlinear behaviour. Memristors are {typically} characterised by a "pinched hysteresis loop" in their current-voltage ($I$-$V$) graph, which intersects at zero voltage and current \cite{Strukov2008}.

A general description of a memristive device is:
\begin{equation}
\frac{dx}{dt} = F(x, u), \quad y = H(x, u)u,
\end{equation}
where $x$ represents internal (physical) states, $u$ is the input ($V$ or $I$), and $y$ the output. The function $H(x, u)$ corresponds to the memristance or memductance~\cite{Caravelli2018}.

The concept dates back to the 19th century with studies on arc lamps and early radio components. However, in 1971, Chua formalised the memristor as the fourth fundamental circuit element, defining it as a device that ``remembers" the history of past inputs.  
From a practical perspective, it is useful to think of a memristive device as a resistance (e.g., satisfying Ohm's law $RI=V$) with a time dependent $R(t)$, whose value depends on the history of $V(t)$ (or $I(t)$, depending on the type of device). In practice, memristive devices can be experimentally realised by exploiting the physical mechanism of resistive switching in nanoscale devices \cite{wang2020resistive}. Key classes of resistive switching memory (ReRAM) devices include \textit{i)} electrochemical metallisation memories (ECM), where metal cations form/dissolve conductive filaments, \textit{ii)} valence-change memories (VCM) where anion/vacancy motion modulates interfacial barriers and filaments \cite{WaserAono2007,Valov2011}, and \textit{iii)} phase-change memories based on reversible amorphous-crystalline structural transitions \cite{WuttigYamada2007}. While variability and stochastic effects typical of experimental devices can be limiting for memory applications, they can also be exploited as a computational resource in neuromorphic devices \cite{Yang2013}. The extension of the two-terminal memristive concept to generic multiterminal memristive systems, as SOMNs, is discussed in Ref.~\cite{milano2026memristance}
\end{tcolorbox}
\end{figure*}
%%%

Beyond scientific interest, SOMNs have the potential to address practical challenges in embedded edge intelligence~\cite{Zhou_edge_2019}. Their ability to learn \textit{in situ} under resource constraints makes them ideal for applications in robotics {and autonomous systems}, as well as for continuous real-time dynamic sensing and predictive analytics (cf. Fig.~\ref{fig:evo}~(b)).
{Indeed, in these edge applications, SOMNs offer a viable solution to AI's energy crisis~\cite{Bourzac2024}.
SOMNs not only bypass the von Neumann bottleneck, by integrating memory and processing in a single system, but also enable efficient learning directly at the physical interface with reduced software complexity.}
To understand how SOMNs can bridge material properties and learning, we examine the physical principles that underpin their brain-like dynamics and highlight how these systems have been implemented to date to perform meaningful learning tasks.

The remainder of this Perspective is organised as follows.
Sec. \ref{sec:dynamics} discusses the brain--like dynamics of SOMNs, including the underlying transport physics that produces physical {plasticity} (sec.~\ref{sec:plasticity}) and criticality (sec.~\ref{sec:crit1}).
Sec.~\ref{sec:theory} presents  a theoretical explication of SOMNs as complex physical systems, including theoretical methods for elucidating how memristive nonlinearity and memory combined with the physical constraints of Kirchhoff's circuit laws induce highly non-trivial dynamics (sec.~\ref{sec:theory1}), and how a mean-field model can be constructed (sec.~\ref{sec:theory2}).
Sec.~\ref{sec:physlearning} discusses two main approaches to {physical} learning demonstrated with SOMNs: physical reservoir computing (sec.~\ref{sec:physrescomp}) and associative learning (sec.~\ref{sec:physasslearn}).
We conclude with an outlook in sec. \ref{sec:outlook}.

%%%%%

\section{Brain-like dynamics in SOMNs}
\label{sec:dynamics}
{SOMNs are multiterminal memristive systems (Fig.~\ref{fig:switching}(a)) where brain-like dynamics arise from nanoscale transport dynamics and network connectivity. Emergent brain-like properties include plasticity, e.g. the ability to adapt to changing input signals; and avalanche criticality, e.g. signal propagation exhibiting long-range spatio-temporal correlations}~\cite{DiazAlvarez2019,Vahl2024}.
Although the underlying nanoscale transport processes in NW and NP networks differ, both types of SOMNs show similar brain-like dynamics related to collective memristive switching effects.

\subsection{Physics of ionic/electronic transport}
\label{sec:transport}

\textit{NW networks.} The dynamics of NW networks are governed by ionic transport in the junctions between NWs. 
 The NWs are comprised of an electromigrating metal and a dielectric and assemble into a random arrangement where a nanoscale electrical junction forms at each contact region at the intersection  between two NWs ~\cite{Milano2019,Kuncic2021}. {Most commonly, NW networks employ silver as the electromigrating metal with either a polymer coating (e.g., Polyvinylpyrrolidone, PVP) \cite{manning2018emergence,DiazAlvarez2019,milano2020brain} and/or metal chalcogenide (e.g., Ag\textsubscript{2}S \cite{Stieg2011,sillin2013theoretical}), though other {materials} have also been explored {(e.g. Ag-TiO$_2$~\cite{Li_etal_2020}, AgI~\cite{Lilak2021}, Ag$_2$Se~\cite{zhu2023online,Kotooka2024thermally})}.} {Under externally applied electrical stimulation,} the strong electric field in nanoscale junctions {can} force Ag\textsuperscript{+} cations to electromigrate towards the opposite wire, where they nucleate back to metallic Ag, and form conductive nanofilaments across the junction \cite{milano2020brain}.
Evidence that the transport is localised into atomic-scale metallic filaments is given by quantum conductance effects, where conductance states are observed in multiples of the fundamental quantum of conductance $G\textsubscript{0}= 2e\textsuperscript{2}/h$ \cite{manning2018emergence, milano2024electrochemical}.

Counter to this additive process, minimisation of the interfacial energy and local electromotive force effects (nanobattery effect) act to dissolve the nanofilaments \cite{valov2013nanobatteries}, becoming the dominant process when the applied voltage is removed. Therefore, similarly to biological neurons where the synaptic strength is associated with changes in ionic flow and to membrane depolarisation and spontaneous repolarisation, NW junctions modulate their conductance depending on input history and are characterised by a short-term memory ranging from milliseconds to seconds. This mechanism is one of the most common memristive behaviours, at the basis of electrochemical metallisation (ECM) memories and atomic switch devices \cite{valov2011electrochemical}. In addition to the external electrical bias, the main intrinsic factors are ion mobility and redox rates \cite{yang2014electrochemical}. 

\begin{figure*}[ht!]
    \centering
\includegraphics[width=0.7\linewidth]{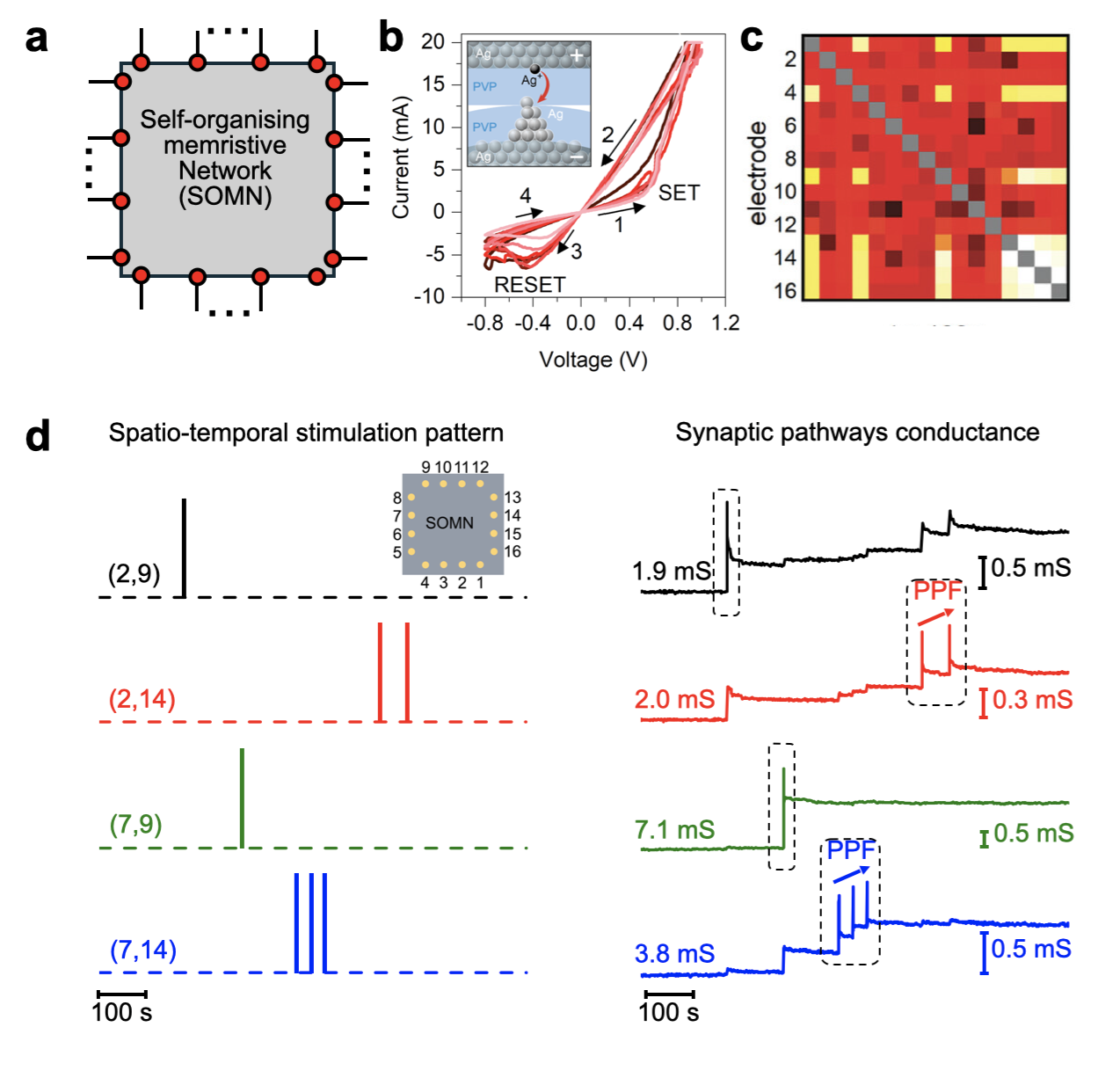}
    \caption{Memristive and brain--like dynamics of SOMNs. (a) Schematic of a multiterminal SOMN device. (b) Example of memristive behaviour of a NW network measured under multiple AC cycles in a two-terminal configuration, showing hysteresis loops in the \textit{I-V} plane typical of memristive systems; line colours (from brown to red) depict cycle number; the inset shows a schematic of the electrochemical metallisation mechanism underlying memristive switching at each junction. Adapted from Ref.~ \cite{milano2020brain}. (c) Conductance matrix of a multiterminal NW-based SOMN device where each matrix element represents the effective resistance between a pair of terminals. Adapted from Ref.~\cite{pilati2024emerging}. (d) Example of synaptic plasticity responses in multiterminal NW-based SOMNs where a spatiotemporal input (left panel, with multiterminal SOMN schematic as inset) drives changes in synaptic conductance (right panel); circled areas highlight changes related to direct stimulation of the corresponding conductance pathway, where temporally correlated inputs results in paired pulse facilitation (PPF). Adapted from Ref.~\cite{milano2023tomography}. }
    \label{fig:switching}
\end{figure*}

Zooming out at the macroscopic level, NW networks are characterised by metastable states induced by the collective interactions among many memristive junctions where nanofilaments are continuously reshaped due to the local redistributions of electric potential and current. Under AC voltage sweeps, NW networks exhibit a collective hysteresis response characteristic of memristive {systems} (cf. Fig.~\ref{fig:switching}b).
When large external biases are applied, or local inhomogeneities are present, Joule heating can induce local breakdown in single nanowires, creating gaps where metal ions electromigrate and form nanofilaments, as in NW junctions \cite{milano2024electrochemical}. While the short-term dynamics of such new memristive elements are likely indistinguishable from those of NW junctions (from the electrical point of view), the network topology, along with all local redistributions of electrical potential and current, is irreversibly changed, thereby imparting long-term memory characteristics to the network. Borrowing terminology from connectomics \cite{sporn1,sporn2}, resistive switching in NW junctions and the creation of new memristive connections across single NWs can be associated with two forms of plasticity: reweighting, where an existing synapse (junction) changes its conductance, and rewiring, where a new synapse is created by forming a previously absent connection across a gap \cite{milano2020brain}.
The collective spatiotemporal dynamics of multiterminal SOMNs can be experimentally investigated by monitoring the evolution of the conductance matrix of the system where each matrix element is the effective resistance between two terminals of the SOMN (cf. Fig.~\ref{fig:switching}c) and/or through voltage maps that provides information about the spatial distribution of the voltage drop when a stimulation is provided. \cite{pilati2024emerging}

\textit{NP networks.}
NP
networks are typically fabricated in vacuum using {nanoscluster production systems that deposit} the NPs on a solid substrate. There, they form a 2D layer of connected groups of particles separated by nanogaps\cite{Sattar2013, bose2017stable}.
With this method, the surface coverage can be precisely tuned to poise the system near the percolation threshold. This is vital as it results in scale-free network structures and consequently scale-free dynamics \cite{Shirai2020}. Interestingly, it has been found that the architectures of 3D percolating networks are similar to those of 2D systems \cite{Bones2025}.
 Typical metallic materials employed are Sn\cite{Sattar2013, bose2017stable, Le2020, Mallinson2019, Mallinson2024}, Bi\cite{Schulze2003}, Au\cite{Minnai2017, Minnai2018, Bose2022, Profumo2025},  Ag \cite{Carstens2022, Rao2023, Gronenberg2024, Pal2025, Pal_nanos_2025, Adejube_2025}, and Cu \cite{vanderRee2025}, but recent work in which devices were fabricated using Mo NPs shows the diversity of materials that can be used \cite{vanderRee2024}. The most recent work on Ag devices \cite{Pal_nanos_2025, Adejube_2025} provides a possible connection to work on NW network devices and suggests that the evolution of Ag network morphology strongly affects memristive switching behaviour.
 
The dominant microscopic conduction mechanisms in NP networks can differ from those in NW networks. In the most widely studied NP systems, notably Sn NP networks measured in vacuum, there is currently no evidence that electrochemical ionic transport plays a role \cite{bose2017stable,Mallinson2019}; instead, conductance is generally consistent with field-driven atomic rearrangements in nanogaps that modulate tunnelling and can evolve into filamentary bridges under stronger driving. Au discontinuous films exhibit closely related field-driven nanogap/filament dynamics both in air and in vacuum \cite{Bose2022}.  These physical mechanisms are in contrast to electrochemical mechanisms that are responsible for the similar effects with NW-based SOMNs \cite{Carstens2022}.
For example, quantum conductance effects have also been observed in NP-based SOMNs \cite{Sattar2013}.

\begin{figure*}[ht!]
    \centering
\includegraphics[width=0.95\linewidth]{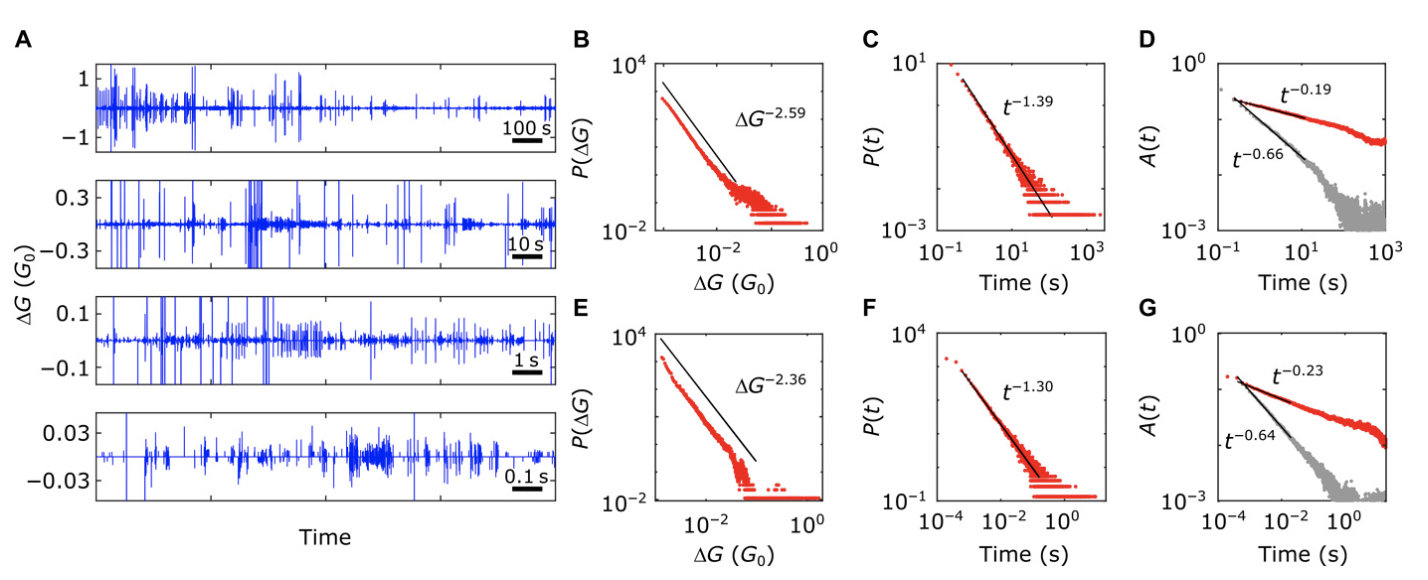}
   \caption{\textit{Scale-free collective dynamics of SOMNs.} (a) Self-similar conductance changes in memristive switching events in NP networks. Top panel: 2400~s of experimental data; lower panels: successive segments of the data with 10, 100 and 1000 greater temporal magnification and with 3, 9 and 27 times greater magnification on the vertical scale (conductance changes shown in units of $G_0=2e^2/h$ for convenience). These avalanches of activity appear qualitatively similar across multiple time scales. {(b,e) The probability density function $P(\Delta G)$ for changes in total network conductance exhibits heavy-tailed distributions. (c,f) Inter-event intervals exhibit power-law statistics, suggesting temporal correlations between events. (d,g) The autocorrelation function of the switching activity (red) decays as a power law over several decades; shuffling the IEI sequence (gray) destroys correlations and yields a steeper decay. Data in (b--d) are measured using a slow sampling rate, whereas (e--g) are measured using a sampling rate which is $10^3$ times faster, providing further evidence for self-similarity.} Adapted from Ref.~\cite{Mallinson2019}, where it was also shown that the avalanches satisfy strict criteria for criticality. Subsequent work demonstrated that power law behaviour was still observed with $\mu$s sampling times \cite{Acharya2021}.}
    \label{fig:avalanche}
\end{figure*}

The memristive mechanism {in NP networks} is described as follows: under an external electrical bias, the strong electric field in the gaps between NPs causes surface atoms to move \cite{olsen2012surface}.
At low electric fields, this force creates ``hillocks" on the surface of NPs that reduce the gap size, thus modulating the tunneling current \cite{mallinson2023reservoir}. At high electric fields (or if the field is applied for a long time), the hillocks extend all the way across the gaps and form filaments \cite{Sattar2013}. Once the bias is removed, the hillocks relax back due to interfacial energy minimisation, {while filaments break due to electromigration effects \cite{Sattar2013}}. The filament formation and breaking processes can be characterised as neuron-like integrate-and-fire mechanisms\cite{Pike2020} and timescales are typically in the microseconds range \cite{Acharya2021}. Filament formation / breaking leads to well-defined changes in conductance i.e., switching events. When one gap switches, it causes a redistribution of voltage and current across the network, leading to avalanches of events that resemble those observed in biological neural networks \cite{Srinivasa2015,Munoz2018} (cf. sec.~\ref{sec:crit1}).

These field-driven morphological changes provide a natural internal state for each junction, with bias-induced growth narrowing the tunnelling barrier and post-bias relaxation widening it again. The junction conductance therefore reflects a history-dependent effective gap size. Because the conductance of NP junctions depends exponentially on the instantaneous tunnelling barrier width (i.e. on the effective gap size), even a modest, history-dependent change in protrusion height produces a strong nonlinear potentiation during stimulation, followed by a gradual decay after stimulation.
It has been shown that the low field hillock and high field filament modes are useful for different kinds of computation \cite{Mallinson_2023, mallinson2023reservoir, Mallinson2024,Heywood2024,Studholme2023,Studholme2024, Studholme2025}.

Charge migration and/or atomic rearrangement in the nanoscale junctions between NWs or NPs are responsible for the memristive switching dynamics in SOMNs. Together with their network connectivity, these mechanisms result in self-adaptation to changes in the input signals (synapse-like plasticity) and critical dynamics (neuron-like avalanches), which are discussed in more detail below.

\subsection{Plasticity}\label{sec:plasticity}

Plasticity refers to the capacity of a system to modulate its internal states or connectivity in response to external stimuli.
In biological neuronal networks, plasticity is fundamental to learning and memory, enabling organisms to retain, discard, and process information in response to new signals.
Following experimental demonstrations of memristive devices as synaptic elements in neuromorphic circuits \cite{jo2010nanoscale, ohno2011short}, a wide range of synaptic functionalities have been realised in SOMNs due to their ability to dynamically reconfigure in response to external electrical stimuli, particularly in multiterminal configurations.

\textit{A brief taxonomy}. 
To facilitate the discussion below, we introduce here the main forms of plasticity typical of biological systems that can be implemented in SOMNs. \textit{Homosynaptic plasticity:} this is a stimulus-specific modulation of synaptic weight (input-specific ``synaptic'' strengthening/weakening). \textit{Heterosynaptic plasticity:} a form of synaptic plasticity where the strength of a synapse changes even though that specific synapse was not directly activated. \textit{Short-term plasticity (STP):}  a temporary change in synaptic strength caused by recent presynaptic activity, producing a fading-memory trace of recent inputs.
\textit{Metaplasticity:} plasticity that depends on the prior history of modifications (past activity changes the \emph{future} plastic response).
\textit{Structural plasticity:} refers to persistent topological changes (creation/removal of conductive connections) that encode longer-term memory.

\textit{Homosynaptic and heterosynaptic plasticity in multiterminal SOMNs.} Experimental results show that the emergent behaviour of NW networks can emulate input-specific changes in synaptic conductance pathways, resembling \textit{homosynaptic plasticity} \cite{milano2020brain,Kuncic_etal_2020}. Additionally, multiterminal memristive devices demonstrate that plasticity can also extend to synaptic pathways not directly stimulated, effectively mimicking \textit{heterosynaptic plasticity} effects \cite{milano2020brain, montano2022grid}. {Heterosynaptic plasticity effects in SOMNs are related to the modulation of not-directly stimulated conductive pathways due to a network-level redistribution of currents/voltages under Kirchhoff constraints.}

\textit{Short-term plasticity and fading memory (volatile internal state).}
These networks also exhibit \textit{short-term plasticity}, where synaptic weights---represented by the {effective} conductance between neuron-like terminals---depend on recent stimulation history, as evidenced by potentiation during stimulation followed by spontaneous relaxation after the end of stimulation. This effect {is related to} volatile memristive mechanisms inherent in network elements \cite{milano2020brain, diaz2019emergent, zhu2021information}. Short-term memory effects, experimentally demonstrated to be controllable across multiple timescales \cite{milano2020brain}, enable the emulation of working memory processes observed in biological brains. Working memory includes not only the temporary retention of information but also its active manipulation \cite{diaz2019emergent,loeffler2023neuromorphic}. Crucially, these short-term, relaxing conductance traces provide the physical basis of a \emph{fading-memory} kernel, which facilitates the implementation of reservoir computing paradigms (sec.~\ref{sec:physrescomp}).

In the case of NPs, this plasticity/memory arises from the same field-driven nanogap dynamics described in Sec.~\ref{sec:transport}: an applied bias drives surface-atom migration that builds a nanoscale protrusion (``hillock'') into an inter-particle gap, thereby narrowing the effective tunnelling barrier and potentiating the junction conductance during stimulation; once the bias is removed, surface-energy-driven relaxation competes with the prior growth and retracts the protrusion, producing a gradual conductance decay that encodes recent input history (a volatile internal state and fading-memory short-term plasticity) \cite{Daniels_2023,Daniels2022}. Note that the intrinsic timescales of these effects ($\mu$s) are much faster than those observed in NW-based SOMNs and plasticity in these systems needs further study. In fact, recent work on PNNs had focused on the incorporation of molecular memristors/synapses into the networks to achieve long term memory effects alongside neuron-like dynamics; complex network dynamics result from the interaction (via Kirchhoff's laws, of the multiple hillocks that are distributed across the network, and can be exploited for RC \cite{Mallinson2023,Monaghan2025,Heywood2025}. 

\textit{Metaplasticity, long-term and structural plasticity (persistent memory).}
Beyond short-term effects, NW networks can also preserve {memory of} previously established pathways while forming new connections. This capability mirrors \textit{metaplasticity} in biological systems, where synaptic plasticity depends on the historical context of prior modifications \cite{loeffler2023neuromorphic}. Furthermore, changes in the NW network's physical topology, known as \textit{structural plasticity}, have been observed through electrically induced rupture and electrochemical rewiring of NWs \cite{milano2020brain, milano2024electrochemical}. These long-lasting changes in SOMNs can be exploited to emulate long-term plasticity effects, a crucial aspect for storing knowledge in the system during learning. In this context, tomographic measurements have experimentally shown that spatiotemporal stimulation patterns in multiterminal memristive NW networks can induce adaptive changes in the effective synaptic conductance of the pathways connecting the stimulated terminals, as illustrated in Fig.~\ref{fig:switching}d. These spatially correlated short-term plasticity effects can further evolve into long-lasting memory traces resembling biological \textit{memory engrams}, namely the physicochemical modifications thought to underlie stored experiences in the brain \cite{milano2023tomography}. Similarly, short-term and long-term memory functionalities have also been reported in NP networks \cite{Srikimkaew2024}. Collectively, these findings demonstrate the potential of SOMNs as substrates for both processing and storing information at the material level.

In particular, it is possible to observe that: \textit{i)} stimulation of a pair of electrodes results in potentiation of the corresponding synaptic pathway, followed by spontaneous relaxation due to volatile effects; \textit{ii)} temporally correlated inputs produce a gradual increase in the effective synaptic conductance, emulating paired-pulse facilitation (PPF); and \textit{iii)} stimulation of a pair of electrodes also gives rise to changes in non-directly stimulated synaptic pathways (heterosynaptic plasticity) due to the connectivity of the system.

\subsection{Criticality}\label{sec:crit1}

Critical phenomena play a fundamental role in various physical systems, characterised by divergent spatio-temporal correlations and the breakdown of mean-field theory (MFT) approximations. Classic equilibrium examples include the paramagnetic--ferromagnetic phase transition described by the Ising model. Importantly, similar critical behaviours emerge in non-equilibrium biological neuronal networks~\cite{Wilting_2019}, where they are believed to underpin efficient information processing.

NW-based SOMNs, which are generally fabricated above the percolation density threshold, exhibit criticality when networks are electrically stimulated just above a critical applied voltage~\cite{Hochstetter2021, Dunham_2021}. 
Avalanche criticality observed in Ag-based NW networks is due to electrical percolation when the network dynamics is dominated by electron tunnelling in the junctions~\cite{Fairfield_2016, DiazSchneider2024}.
This occurs when junction tunnel gaps form as Ag filaments evolve. These gaps continuously close and re-open as filaments grow and stochastically decay, driving rapid memristive switching between metastable conductance states.  
At voltages well above the critical voltage, some memristive junctions can form stable filaments that support ballistic transport. When this occurs, NW networks undergo a discontinuous dynamical phase transition resembling dielectric breakdown (also referred to as a ``bolt phase''), characterised by the emergence of large amplitude avalanche events~\cite{Shekhawat2011, Sheldon2017, Hochstetter2021, DiazSchneider2024}.

The first evidence of avalanche criticality in SOMNs was demonstrated in NP networks~\cite{mallinson2019avalanches}, which are fabricated at the percolation threshold and thus exhibit scale-free responses to electrical stimulation. As shown in Fig.~\ref{fig:avalanche}, conductance activity patterns (avalanches of switching events) are self-similar across multiple time scales and can be analysed through the probability density function, the inter-event intervals statistics and autocorrelation function.

Recent experiments \cite{michieletti2025self} provide deeper insights into critical phenomena in NW networks. These studies demonstrate the programmable induction of critical states via targeted electrical stimulation. Importantly, spatially distinct regions within NW networks can exhibit and sustain local critical dynamics, facilitating spatially correlated interactions that directly influence computational functionality.

While both NW and NP networks show critical dynamics, their measured exponents deviate from known universality classes, hinting at novel behaviour.
This may arise due to higher--order interactions or specialised percolation dynamics not fully captured by standard theoretical percolation models~\cite{Sun2023, Caravelli2016}, as previously suggested~\cite{Fostner2014, Shirai2020, Daniels_2023}. In particular, the combination of tunneling transport and memristive switching dynamics may distinguish SOMNs from classical percolation universality classes.
Recent studies show that tuning network dynamics, via global feedback \cite{Scharnhorst_2018} or local control of junctions \cite{michieletti2025self}, can steer systems toward different universality classes, offering pathways to control and apply criticality in SOMNs.

Avalanche criticality in SOMNs is an important concept, as it is a prominent feature that mirrors experiments performed in neuroscience \cite{Beggs2003}. It is in fact hypothesised that functioning near this point optimises information processing \cite{Srinivasa2015, Munoz2018, Wilting_2019, hengen_criticality_2025}. {SOMNs are} controllable and experimentally accessible physical systems, and offer valuable abiotic platforms for rigorously testing this hypothesis through direct implementation and analysis of computational tasks (discussed further in sec.~\ref{sec:physlearning}).

%%%%%

\section{SOMNs as dynamical complex systems}
\label{sec:theory}

As highlighted in the previous section, SOMNs are complex and dynamic electrical circuits that produce emergent nonlinear dynamics. Here, we present a theoretical framework that explicitly treats SOMNs as dynamical complex systems subject to the physical constraints of Kirchhoff's electrical conservation laws. This theoretical framework enables models to be developed to gain deeper insights into how the experimentally observable collective dynamics arise from memristive interactions.
In particular, this framework is useful for identifying dynamical transitions and to characterise the dynamical regimes that may enhance learning~\cite{Sheldon2022,Baccetti2024,Barrows2025unc}, as elaborated in sec.~\ref{sec:physlearning}. Sec.~\ref{sec:theory1} presents a theoretical framework of SOMNs in terms internal memory parameters, while sec.~\ref{sec:theory2} presents a macroscopic mean-field description taking into account Kirchhoff's laws and its relation to experimentally observed collective properties. 

\subsection{Theoretical Methods}\label{sec:theory1}

A practical theoretical framework for modelling SOMNs is the lumped-circuit approximation (i.e. an approximation of the underlying physics in terms of electrical elements and Kirchhoff's laws), {which mathematically describes SOMN networks through a graph-theoretical approach}. 
By representing their conductance dynamics with memristive elements and voltage drops across the junctions, SOMNs can be effectively modelled as assemblies of interacting subcircuits, each corresponding to an individual memristive junction (neglecting the resistance of individual NWs or NPs). Their interactions are governed by Kirchhoff's laws and constrained by their underlying circuit topology, resulting in complex, high-dimensional nonlinear dynamical systems.

A convenient way to mathematically capture the diverse nonlinear dynamics observed experimentally is to model each junction's conductance as a dynamical variable dependent on its voltage or current history. This approach provides a simplified model that captures memory and adaptation using only voltage, current, and time, i.e., physical observables.
This method applies to both NW \cite{Caravelli_2023c} and NP \cite{daniels2022reservoir} networks and can be extended to other systems \cite{Barrows2025b}.

The internal state of each memristive junction typically evolves according to a first-order differential equation of the general form (cf. \textbf{Box~1}):
\begin{equation}
	\frac{dg}{dt} = F\bigl(g,\,\delta v(t),\,t\bigr), \quad G = G(g),\label{eq:nanojunc}
\end{equation}
where $g \in [0,1]$ represents a normalized internal memory parameter, and $v(t)$ denotes the local instantaneous voltage across the junction. The function $F$ encodes how the conductance changes in response not only to $v(t)$ but also to make explicit time dependence, allowing modelling of more complex memory behaviours. $G(g)$ represents the electrical conductance of the junction in terms of $g$.
An equivalent representation can also be made in terms of electrical resistance.

{While the lumped-element model introduced above is widely applicable,
alternative modelling approaches have been developed, with different memristive mechanisms. For instance, a potentiation--depression rate balance equation, originally derived from detailed studies of single memristive junctions~\cite{miranda2020modeling}, has been used to describe dynamic conductance changes at both the junction level~\cite{milano2022connectome} and in spatially extended NW networks via a grid-graph modelling framework~\cite{montano2022grid}. These rate-based models offer a complementary theoretical perspective by naturally capturing key plasticity phenomena, including short-term plasticity, metaplasticity, and structural plasticity, as emergent from cumulative electrical activity.

The lumped-element approximation simplifies the system by reducing each junction to a single memory variable, omitting other physical degrees of freedom that may affect network dynamics. For instance, small but finite capacitive effects at the junctions, though often excluded from basic memristive models, have been experimentally observed and may become significant under certain conditions \cite{Merle2022, Merle2023}. These effects introduce distinct memory-like dynamics and can be examined using techniques such as Nyquist impedance spectroscopy \cite{Lazanas2023,Merle2023}.

From a practical standpoint, Kirchhoff's laws can be numerically solved for a given network and initial junction states to determine voltages and currents across each element. Integrating the dynamical equations (\ref{eq:nanojunc}) then reveals the time evolution of the conductances. This direct approach is widely used in the literature \cite{Kuncic_etal_2020, zhu2021information, Pike2020, daniels2022reservoir,milano2022connectome,Hochstetter2021, mallinson2023reservoir} to assess the effective conductance and functional behaviour of SOMNs.

To expose how topology and constraints shape dynamics and to gain analytical insight, 
an {analytical} method using the projector operator formalism (see ~\textbf{Box~2}) has been developed\cite{Caravelli_2017b}. This incorporates Kirchhoff's constraints into the junction evolution equations via projection matrices: $\Omega_B$ in the nodal formalism and $\Omega_A$ in the mesh formalism.
Within this framework, the network dynamics reduce to a system of coupled nonlinear ordinary differential equations:
\begin{equation}
	\frac{d\vec g}{dt} = \vec{\mathcal F}\bigl(\vec g,\,\vec{\delta v(t)},\,\Omega_B,\,t\bigr),\label{eq:nanojuncnet}
\end{equation}
where $\Omega_B$ ensures explicit enforcement of Kirchhoff's laws.

In equations like (\ref{eq:nanojuncnet}), nonlinear memory effects emerge from the interplay between circuit topology and time-dependent junction conductances. Unlike resistive networks with fixed internal states, memristive networks continuously adapt their conductance (or resistance) in response to external driving. The topology is captured via projector matrices, $\Omega$, which typically appear in the equations when Kirchhoff's laws are implemented analytically 
also via matrix inverses of the form $(I \pm \chi \Omega \mathcal{G})^{-1}$, where $\mathcal{G}_{ii} = g_i$ is a diagonal matrix \cite{Caravelli_2017b,Caravelli_2023c}.

This formalism can be extended to more general network motifs, where edges represent subcircuits with more complex, higher-order behaviour, such as inductive or capacitive coupling, in addition to memristive \cite{Barrows2025}. Though not explored in detail here, such extensions may be important for accurately modelling phenomena like the experimentally observed capacitive effects in SOMNs \cite{Merle2022,Merle2023}.
Crucially, the matrix inverse structure in the projector formalism induces nontrivial feedback loops within the network. These loops underpin emergent behaviours such as
%hysteresis,
self-organisation and dynamical phase transitions \cite{Caravelli_2023c}, providing a unified analytical framework for linking local memristive dynamics to rich global behaviour.

Incorporating stochasticity into SOMN dynamics offers a powerful extension of the theoretical framework. 
Stochasticity is a fundamental characteristic of SOMNs~\cite{Pike2020, milano2025self}, particularly in the high-voltage regime where the formation and rupture of conductive filaments leads to continuous redistributions of voltage and current, thereby triggering new switching events. The inherent complexity of these networks~\cite{Shirai2020} further implies that the timing of such events is effectively random. Moreover, the filament formation process itself is intrinsically stochastic~\cite{Acharya2021}. 
Accordingly, SOMNs can be modelled as stochastic dynamical systems~\cite{milano2025self}, with network conductance described, for example, by an Ornstein--Uhlenbeck process that blends stimulus-driven evolution with intrinsic fluctuations. This stochastic mean-field approach captures features such as conductance noise and jumps, while enabling analysis of the potential landscape and the interplay between noise, steady states, and computational capability. In this way, stochastic models provide deeper insight into collective behaviour that cannot be fully described by deterministic frameworks.

The ``solid--liquid brain'' metaphor \cite{Adamatzky2015} helps clarify, at an intuitive level, why plasticity is central to SOMNs. Unlike static resistive networks (``solid brains''), whose connectivity and functional response are essentially fixed, SOMNs behave more like ``liquid brains'': external stimuli continuously reshape their effective conductance pathways (the internal states), allowing the system to adapt its internal state in a way that is closer to biological substrates \cite{Pershin2013,Deng2015}.  These models enable us to describe and even guide the evolution of useful dynamical states, allowing for learning as a form of \emph{guided self-organisation}~\cite{prokopenko}. In this sense, plasticity is not an incidental property, but a defining ingredient of the collective dynamics discussed in this section, because it is precisely the coexistence of memory, reconfigurable connectivity, and Kirchhoff-constrained feedback that gives rise to rich emergent behaviour \cite{Barrows2025b} From a theoretical perspective, equations of the form~(\ref{eq:nanojuncnet}) explicitly combine memory and topology, while also allowing the construction of Lyapunov functions and hence a stability analysis akin to that of Hopfield networks. These functions reveal attractor dynamics in an effective potential landscape \cite{caravelliscience}, showing how stimulus-dependent plastic reconfiguration can generate nonlinear responses, metastability, and exploration of multiple configurations. Such features are directly relevant for optimisation and learning, but they also indicate that SOMNs should be viewed as non-equilibrium complex systems in their own right. For this reason, beyond mean-field and circuit-theoretic approaches, novel theoretical methods drawn from complex systems and disordered systems may prove especially valuable for understanding how plasticity, disorder, and feedback jointly shape their emergent dynamics.

The above exposition provides a phenomenological description of SOMNs via Kirchhoff-constrained dynamical equations for the full ensemble of junction states $\vec g$. Many experimentally accessible properties, however, concern \emph{macroscopic} observables such as the effective conductance $G$ and the emergence of collectively distinct low- and high- conductance regimes under changes in control parameters (e.g. voltage, density, connectivity, noise). This motivates a complementary coarse-graining: in the next subsection, we introduce a mean-field theory (MFT) that reduces the high-dimensional network dynamics to an order-parameter description (e.g., $\langle g\rangle$ or $G$), thereby enabling analytical insight into when conductance transitions occur and where the approximation breaks down near critical regimes.

\subsection{Mean field theory of conductance transitions}\label{sec:theory2}
Critical phenomena in SOMNs occur in at least two related but distinct forms. First, avalanche criticality has been observed experimentally in both NW and NP networks, where the electrical response exhibits scale--free switching statistics and long-range spatiotemporal correlations~\cite{Mallinson2019,Hochstetter2021,Dunham_2021,Pike2020,Bose2022,Acharya2021} (see Fig.~\ref{fig:avalanche}). In NP systems in particular, these behaviours are closely connected to transport near the percolation threshold, so percolation-based descriptions provide a natural starting point for interpreting the emergence of scale--free dynamics. Second, SOMNs appear to exhibit memory-driven conductance transitions that are not purely geometrical, but instead arise from the interplay between memristive internal states, external driving, and Kirchhoff-constrained network feedback~\cite{caravelliscience,Caravelli2016,Hochstetter2021}. In this case, the transition refers to the emergence and coexistence of multiple macroscopic conductance states, and to switching or bifurcation between low- and high-conductance regimes as control parameters such as voltage, density, or connectivity are varied. Analogies to random fuse models~\cite{Zapperi1999}, dielectric breakdown~\cite{Shekhawat2011}, and tunneling-dominated percolation~\cite{Fostner2014} remain useful for framing these behaviours, but they do not fully capture the role of memory in shaping such dynamical transitions. From this perspective, avalanche criticality and percolation are established experimental motifs in SOMNs, whereas these memory-induced conductance transitions are also of theoretical interest in their own right, because they point to a class of non-equilibrium collective phenomena that likely require new analytical and computational methods to be understood fully.

Numerical studies further suggest that these conductance transitions are reflected in changes of the global conductance patterns of the network. To clarify whether such critical points exist, how many macroscopic states can coexist, and what determines the order and universality of the transition, analytical mean-field approaches are particularly valuable. For example, mean-field descriptions of the percolation transition in memristive networks have been developed by approximating each junction as a threshold-based conductance switch~\cite{Sheldon2017}, building on ideas from random fuse models~\cite{Zapperi1999,Fostner2014}. While such approaches capture useful aspects of the collective behaviour, they are not sufficient to predict critical exponents in SOMNs quantitatively, since this requires a more refined treatment of junction-level dynamics~\cite{wilson}, of transport mechanisms such as tunneling, and more generally of the feedback between memory and Kirchhoff-constrained network dynamics. For this reason, these transitions are not only relevant for interpreting experiments, but also constitute a theoretical problem \textit{per se}, motivating the development of novel methods capable of treating memory, disorder, and non-equilibrium feedback on equal footing.

\begin{figure*}[ht!]
	\centering
	\includegraphics[width=0.95\linewidth]{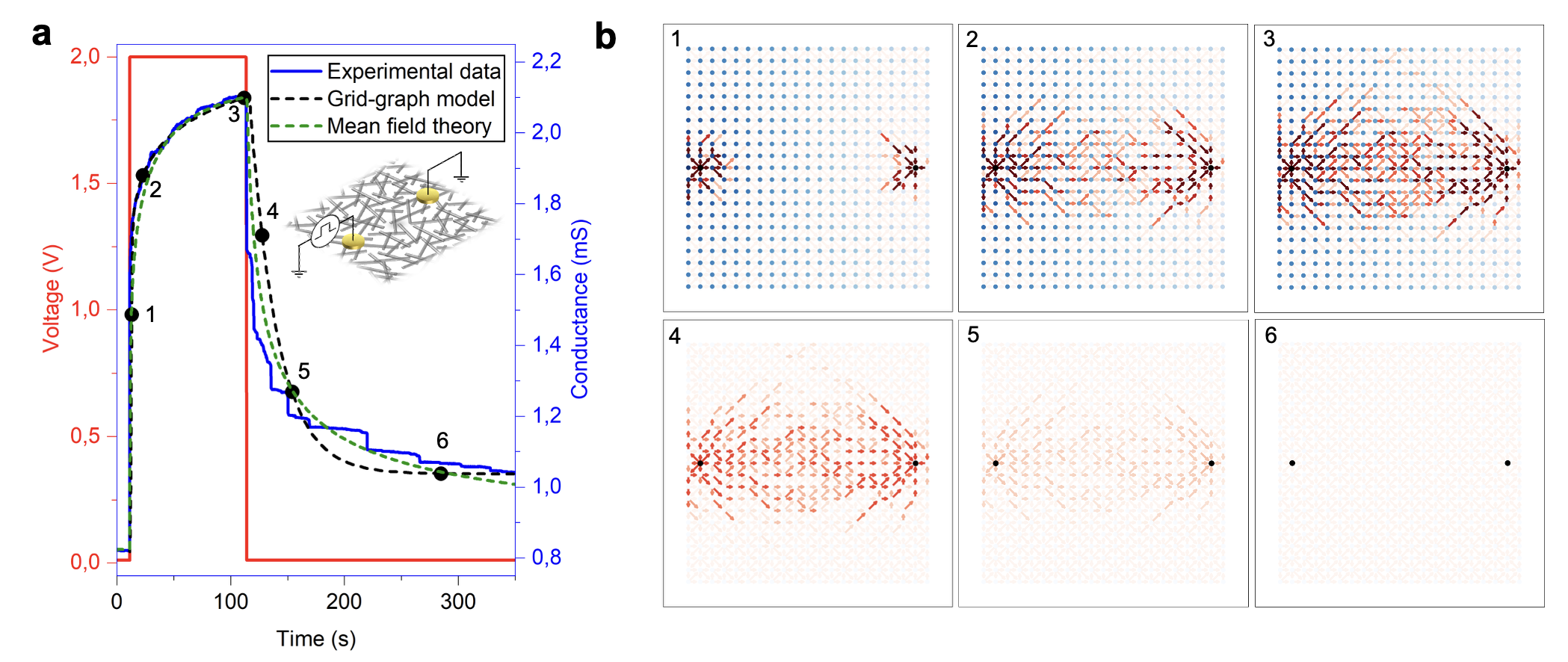}
	\caption{modelling emergent behaviour of memristive networks. (a) Experimental conductance measurements (blue) showing potentiation and relaxation dynamics related to short-term plasticity effects in NW networks under voltage pulse stimulation ({red}), corresponding grid-graph modelling (black dashed line) and mean-field theory results (green dashed line) obtained by using methods outlined in sec. ~\ref{sec:theory}. (b) Corresponding grid-graph modelling visualisation of spatiotemporal network dynamics, characterised by the progressive formation of a conduction path bridging electrodes that progressively fades way after the end of stimulation. Red intensity is proportional to edge conductance, while blue intensity is proportional to node voltage. Adapted from Ref. ~\cite{milano2022materia} and ~\cite{Caravelli_2023c}.}
	\label{fig:mfteffcond}
\end{figure*}

\begin{figure*}
	\begin{tcolorbox}[title=Box 2: From Circuit Topology to Projector Operators, colback=cyan!5!white, colframe=cyan!75!black, width=\textwidth]
		\begin{minipage}[t]{0.48\textwidth}
			To build intuition, a memristive network can be viewed as a dynamic plumbing system: voltages act like pressures, currents like fluid flow. Projector operators then serve as mathematical tools that enforce physical constraints, ensuring flow conservation at junctions. They help capture how electrical signals propagate through the network following Kirchhoff's laws, enabling analysis of dynamic information pathways.
			The projector operator formalism described here for the study of memristive networks is derived from the lumped circuit topology approximation of a memristive network. 
			For a graph-theoretic and more formal introduction to these projectors, see Ref.~ \cite{Zegarac2019,Barrows2025}.\\\ 
			
			The projector operators are of the form
			\begin{equation}
				\Omega_B = B(B^T B)^{-1} B^T,\hspace{1cm} \Omega_A = A(A^T A)^{-1} A^T,
			\end{equation}
			where $\ ^T$ represents the transpose, {$B$ is a circuit's directed incidence matrix and $A$ is its loop matrix,}
			defined below. 
			{Projector operators satisfy $\Omega^2 = \Omega$,}
			implying that the eigenspectrum is only 0's and 1's. \\\ \\
			For a resistive circuit, each branch $j$ (edge) is a resistor with current $i_j$ and $B$ describes how branches connect to vertices (nodes) in the circuit given an edge orientation:
			%which can be given arbitrarily:
			\begin{equation}
				B_{ji} =
				\begin{cases}
					-1 & \text{if edge } j \text{ leaves vertex } i, \\
					+1 & \text{if edge } j \text{ enters vertex } i, \\
					0 & \text{otherwise.}
				\end{cases}\nonumber
			\end{equation}
			$B^t \vec i=0$ then enforces Kirchhoff's Current Law (KCL) on the current vector {$\vec i$}, ensuring current balance at nodes. 
			On the other hand, given arbitrary branches (currents) and mesh orientations (defined as a closed sequence of branches), Kirchhoff's Voltage Law (KVL) can be written in terms of $A$, which represents cycles in the graph, with rows corresponding to loops and columns to branches:
			\begin{equation}
				A_{il} =
				\begin{cases}
					+1 & \text{if edge } i \text{ is oriented alike 
						loop } l, \\
					-1 & \text{if edge } i \text{ is antioriented as 
						loop } l, \\
					0 & \text{otherwise.}
				\end{cases}\nonumber
			\end{equation}
			In terms of this matrix, given the vector of voltage drops $\delta\vec  v$, the KVL can be written as $A^t \delta \vec v=0.$\\\ 
			Thus, KVL and KCL can be written as:			{
				\[
				\mbox{KVL:} \;\; \Omega_A \delta \vec{v} = 0 \qquad , \qquad \mbox{KCL:} \;\; \Omega_B \vec{i} = 0
				\]
			}
			Via cycle and nodal analysis duality arguments (resulting in $B A^T = 0$), it follows that $\Omega_A + \Omega_B = I$,  which shows that $\Omega_A$ and $\Omega_B$ are complementary and project onto cycle and vertex spaces, which can be shown to be equivalent to the voltage and current spaces respectively \cite{Barrows2024}. 
		\end{minipage}
		\hfill
		\begin{minipage}[t]{0.48\textwidth}
			The dynamics of memristive networks can be written directly in terms of these projectors, making Kirchhoff's laws in the dynamics explicit. For instance, for a graph whose edges are composed of resistors $j$ with value $R$ Ohms, and voltage generators in series $v_j$ Volts, the \textit{network} Ohm's law can be written as \cite{Zegarac2019} 
			$R \vec i= \Omega_A \vec v, $ where $i_j$ is the solution of the current on the resistor $j$ at equilibrium. The role of the projectors is to ``network" and integrate the voltage generators to provide voltage drops on the circuit elements.\\\ \\
			{A deep link exists between the graph Laplacian \cite{milano2022connectome} and the projector operators \cite{Barrows2025}. 
			The (weighted) graph Laplacian is defined as \cite{Chung1996}
			\begin{equation}
				L = B^T G B,
				\label{eq:laplacian}
			\end{equation}
			where $G$ is a diagonal matrix of edge conductances (note that in standard circuit theory, $G_{jj} = \frac{1}{R_j}$,}
			with $R_j$ the resistance of edge $j$). For a connected graph, $L$ has rank $n-1$ with $n$ nodes. The pseudoinverse $L^+$ also appears in standard formulas for effective resistances. The quantity most relevant to experiments in two-point conductance measurements is the effective resistance $R_{\mathrm{eff}}(a,b)$ between nodes $a$ and $b$, given by \cite{Klein1993}
			$  R_{\mathrm{eff}}(a,b) = L^+_{a,a} + L^+_{b,b} - L^+_{a,b} - L^+_{b,a}$.
			The non-orthogonal projector $\Omega_{B/R^{-1}}$ enforcing current conservation on edges can be expressed in terms of the Laplacian and its pseudoinverse
			$\Omega_{B/R^{-1}} = B (B^T R^{-1} B)^{-1} B^T R^{-1}= B^T L^{+} B G$, 
			where the second equality uses $L = B G B^T$ and $G = R^{-1}$, and
			where $(\cdot)^{+}$ denotes the Moore--Penrose pseudoinverse. A short calculation shows that $\Omega_{B/R^{-1}}$ is idempotent ($\Omega_{B/R^{-1}}^2 = \Omega_{B/R^{-1}}$), retaining the defining property of a projector. Moreover, $\Omega_{B/R^{-1}}$ can be related to the orthogonal projector $\Omega_{B}$ (see Ref.~\cite{Barrows2024}). \\\ \\
			Both projector operators ($\Omega_{B}$, $\Omega_{B/R^{-1}}$) and the graph Laplacian ($L^+$) encode topological constraints in linear and memristive circuits: $L^+$ captures node-based current conservation, while the projector explicitly enforces it on edge currents. For further discussion on these connections between projection matrices, the Laplacian, and circuit-theoretic graph operators, and properties of planar circuits, see Refs.~\cite{Zegarac2019,Barrows2025,Barrows2025b,Caravelli2017}. {These formal structures are not only rigorous but also essential for analysing how memory and dynamics emerge collectively from simple local rules, laying the groundwork for understanding SOMNs as learning substrates.}
			%\end{cvbox}
		\end{minipage}
	\end{tcolorbox}
\end{figure*}
%%%%%

Using the theoretical framework outlined in sec.~\ref{sec:theory1}, a dynamical MFT has been formulated that attempts to characterise the voltage-induced dynamics of SOMNs by reducing equations of the type (\ref{eq:nanojuncnet}) to a single mean-field dynamical equation. This is written in terms of an effective potential $\mathcal{V}_{\Delta v}(\langle g \rangle)$, which governs the evolution of the network's average conductance state~\cite{caravelliscience,Caravelli_2023c}, and allows determination of whether the interacting system is in the low or high conducting phase. {These phases are determined by the state of the junctions, e.g. all on or all off.}%
\ The resulting equation of motion is of the form:
\begin{equation}
	\frac{d \langle g \rangle}{dt} = -\frac{d}{d \langle g \rangle} \mathcal{V}_{\Delta v}(\langle g \rangle),\label{eq:mfteq}
\end{equation}
where $\Delta v$ is the {voltage drop on the device}. As {applied} voltage increases, the shape of $ \mathcal{V} $ changes, potentially introducing multiple minima corresponding to different {macroscopic} conducting states {corresponding to low- and high- conductance}. A mean-field equation of the type (\ref{eq:mfteq}) has been validated experimentally~\cite{Caravelli_2023c} and Fig. \ref{fig:mfteffcond} shows a comparison between experimental data on short-term synaptic plasticity effects and theoretical response curves obtained through MFT, as well as simulation results based on a grid--graph model that {allows direct visualization of the spatiotemporal dynamics}~\cite{milano2022materia}.

\begin{figure*}[!t]
	\centering
	\includegraphics[width=0.95\linewidth]{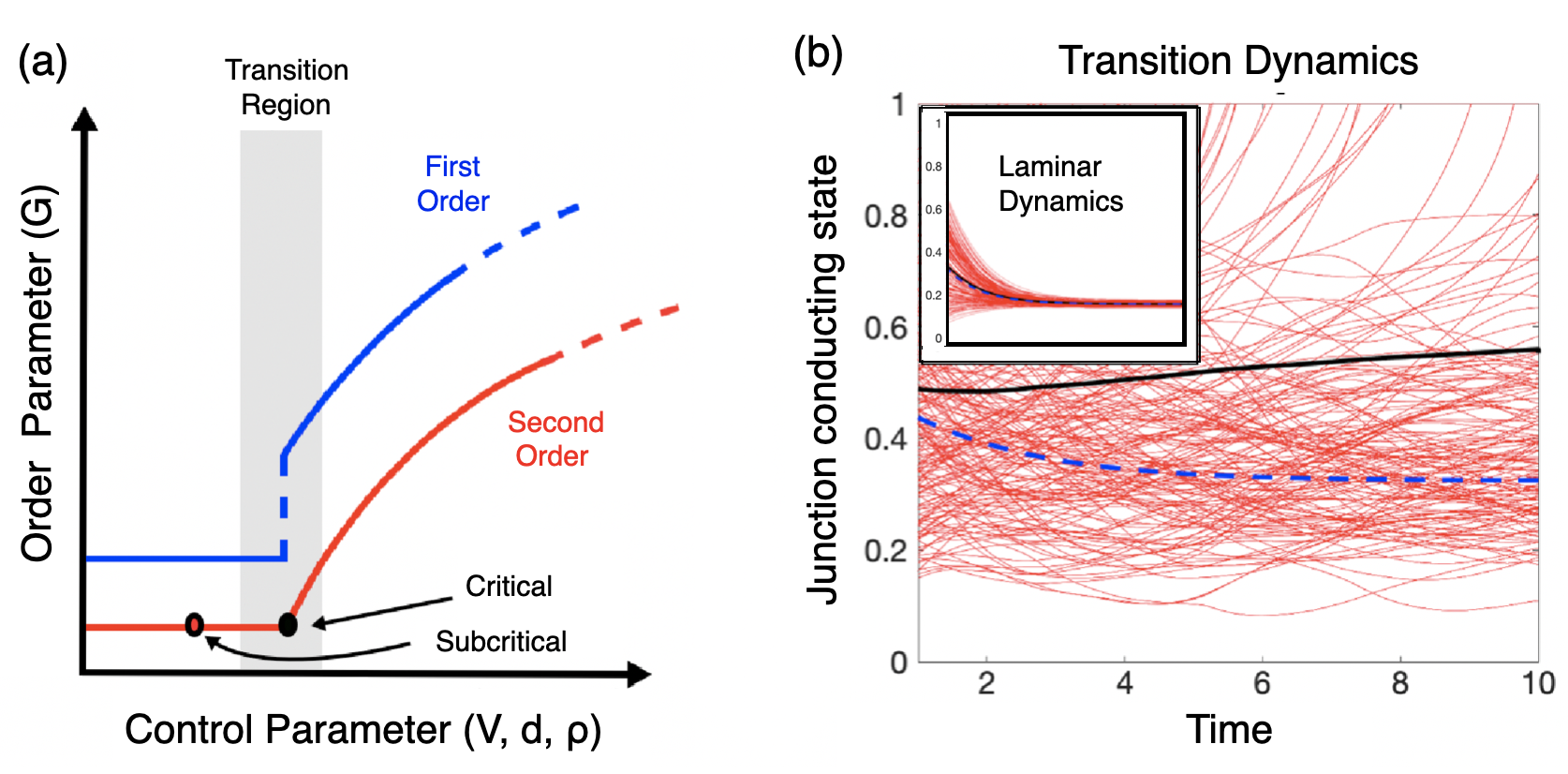}
	\caption{
Conductance transitions in memristive networks.
\textbf{a.} Schematic illustration of conductance phase transitions in SOMNs, where the macroscopic order parameter (average conductance $G$) depends on control parameters such as applied voltage $V$, nanowire/nanoparticle {(areal)} density $\rho$, or average connectivity $d$. The shaded region represents the transition regime where emergent collective behaviour manifests. Depending on system parameters and underlying dynamics, this transition may be continuous (second-order) or abrupt (first-order),
reflecting different types of dynamical behaviours. 
\textbf{b.} Numerical simulation results showing the time evolution of internal junction conductance states (red lines) across a memristive network following a voltage step near the critical point, when the effective potential develops multiple competing minima. The black line shows the ensemble average $\langle g \rangle$ across all junctions, while the dashed blue line corresponds to the prediction from mean-field theory (MFT). The divergence between simulation and MFT near the critical regime illustrates the breakdown of mean-field approximations due to the onset of strong feedback and network reconfiguration. The \textbf{inset} shows the subcritical regime (single minimum in the effective potential), where junction conductances decay smoothly (laminar dynamics) and MFT matches simulations well. These results highlight the emergence of collective, nonlinear dynamics and the importance of including feedback and stochastic effects beyond mean-field approaches. Adapted from Ref.~\cite{caravelliscience}.
}
	\label{fig:transitions0}
\end{figure*}

Various methods have been used to obtain an equation of the form (\ref{eq:mfteq}) from eqn.~(\ref{eq:nanojuncnet}). At the core of these {approximation} methods is how the matrix inverses of the form $ (I \pm \chi \Omega_B \mathcal{G})^{-1}$, representing loosely the nonlinearity induced by Kirchhoff's laws, enters into eqn. (\ref{eq:nanojuncnet}). This matrix inverse, once expanded, leads to nonlinear terms between the $g_i$ variables and thus couples the junctions 
memory states across the network, and its structure gives rise to feedback and dynamical phase transitions in the order parameter $G$. Depending on the shape of the effective potential, these transitions can be both of the first or second order (continuous or discontinuous) type, dependent on voltage, in addition to static geometrical effects due to density. This is depicted schematically in Fig.~\ref{fig:transitions0}(a).
Beyond continuous and discontinuous phase transitions, SOMNs can in principle exhibit more complex transition phenomenology (e.g. higher-order, discontinuous at higher derivatives), depending on microscopic dynamics and network heterogeneity; these are not shown in Fig.~\ref{fig:transitions0}(a) for clarity.

There are various existing theoretical methods to derive a mean field theory. The simplest method for obtaining the mean-field matrix inverse is based on a random matrix theory averaging of the inverse while keeping fixed row and column sums ~\cite{caravelliscience, Bartolucci2023, Bartolucci2024}, corresponding to global information such as the total number of junctions. A more involved method allows preserving certain graph-theoretic properties (e.g., number of cycles), and using averages over random orthogonal transformations $\Omega_B' = O \Omega_B O^T$ (with $O$ an orthogonal matrix), enabling exact averaging over transformations via Weingarten calculus~\cite{Caravelli2024, Collins2022}. Alternatively, and more brutally in \cite{Caravelli_2023c}, the mean-field inverse has been obtained by minimising the Frobenius norm between the exact and the mean-field inverse. The resulting effective potential shape depends on the voltage, which implies that conductance transitions between the low and high conductances are controlled by a macroscopic theory. 
When there is only a single minimum, the dynamics of the junction is laminar {(a relaxational dynamics with negative Lyapunov exponents)} and the MFT works well numerically.
However, near transition points, e.g. when the effective potential develops multiple minima, the system becomes susceptible to noise, which can be amplified and can induce transitions between these two states and a failure of the MFT, as shown in Fig.~\ref{fig:transitions0}(b).

In the transition regime, the Lyapunov exponent associated with the state trajectories becomes positive, implying a turbulent or possibly transiently chaotic regime \cite{tchaos}.
As MFT is not effective near dynamical transitions, alternative approaches (e.g. self-organised criticality~\cite{Bak1996}) and new techniques \cite{Yao2025} are needed to better understand the emergent behaviour of SOMNs near these critical regimes.

Thus, theoretical models provide a foundation for understanding how SOMNs generate the collective nonlinear dynamics observed experimentally.
	From lumped-element to mean-field approaches, these frameworks reveal how memory, phase transitions, and adaptive responses emerge. These {emergent dynamics represent physical information encoding and form the basis of a physical learning} substrate, especially near transitions. The next section explicitly shows how these dynamics enable learning.

%%%%%

\section{Physical learning with SOMNs}
\label{sec:physlearning}
%{I added this introduction}

\begin{figure*}[ht!]
	\centering
	\includegraphics[width=0.9\textwidth]{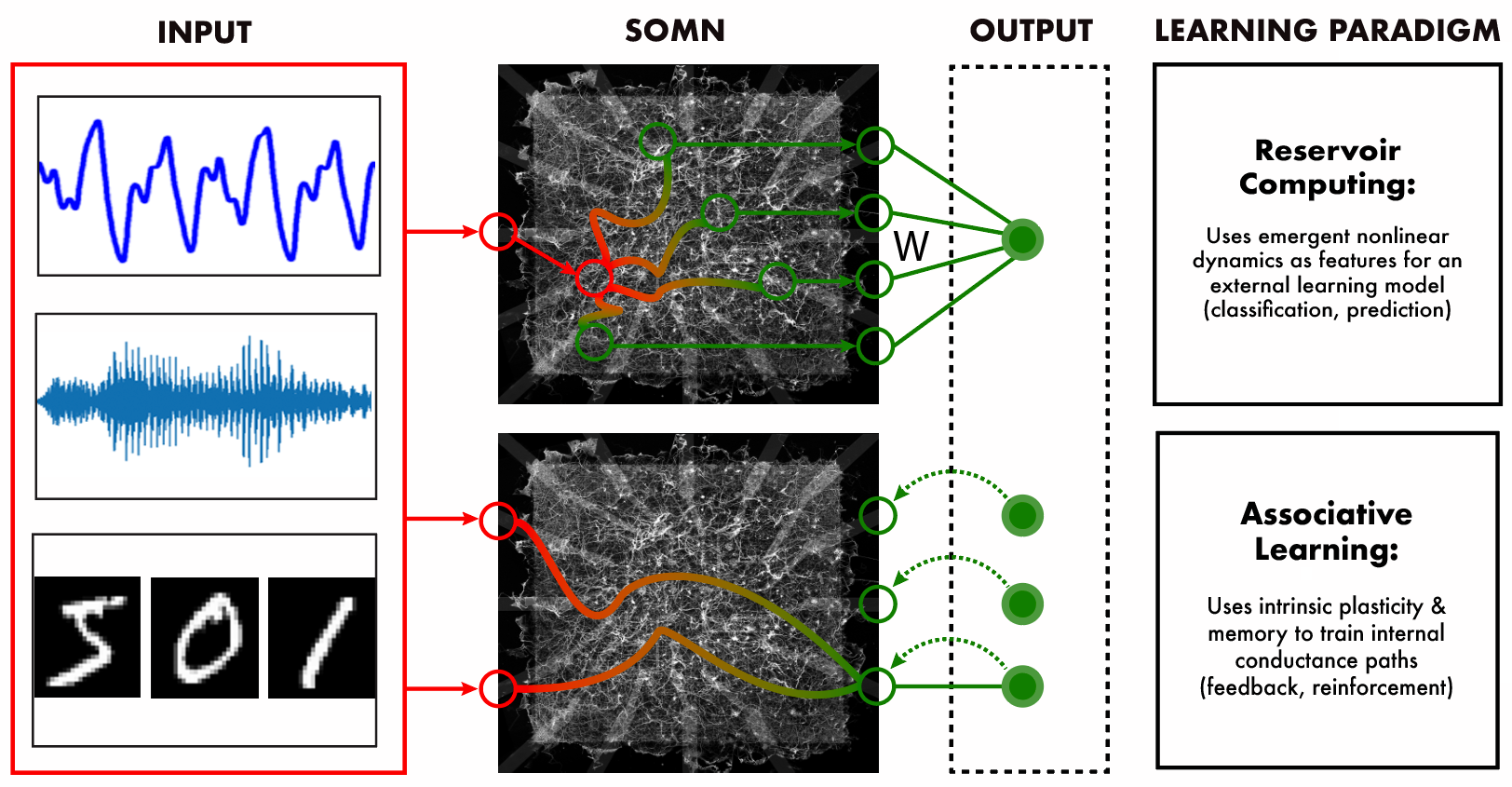}
	\caption{Physical learning paradigms for self-organising memristive networks. Arbitrary input data (left) is converted into voltage signals and delivered to a multi-terminal SOMN device (centre main), where local conductance states of nanoscale memristive junctions adapt to the input signal patterns. Dynamic reconfiguration of the SOMN, governed by internal transport dynamics {(ion migration, atomic filaments/hillocks, electron tunnelling)} and constrained by Kirchhoff's laws in a recurrent network circuitry, performs physical nonlinear transformation of signals to enable learning (output). Two representative learning paradigms are shown (right): in reservoir computing, the SOMN's high-dimensional dynamical features are read out (green solid curves) and exploited in an external machine learning model (weights W);
    in associative learning, feedback is applied locally based on output (green dotted lines); {this can be either at the input or the output or both, via applied voltage changes and electrode selection}.
		Both approaches enable physical substrate-based learning using only input-output access, without any direct addressability of memristive components or global training schemes.}
	\label{fig:Phys-learning}
\end{figure*}

With an understanding of the physical mechanisms that underlie adaptive dynamics in SOMNs, we now show how these properties can be exploited in physical learning systems. To place this in context, it is useful to recall that solving nontrivial learning tasks requires more than simple linear transformations, as became clear after the introduction of the perceptron~\cite{rosenblatt1962principles,minsky2017perceptrons}. In practice, this expressive power can be achieved through multilayer architectures~\cite{cybenko1989approximation}, recurrent dynamics~\cite{hopfield1982neural}, or combinations of both.

Neuromorphic approaches aim to realise such nonlinearities by mimicking biological synapses and/or neurons, either through emulation or simulation~\cite{Christensen_2022}. The emulation approach involves exploiting materials and devices (e.g. memristors) that natively exhibit brain-like nonlinear dynamics~\cite{Mehonic2020}. The simulation approach instead typically involves algorithms that simulate point neuron spike models~\cite{Schuman2022}. While both approaches can mimic synaptic plasticity, which is generally regarded as the key mechanism underlying the brain's unique learning abilities~\cite{schmidgall_brain-inspired_2024}, how this is implemented at the circuit and device level differs considerably. For example, synaptic conductances in memristor crossbar devices are often used to physically instantiate matrix--vector operations to perform inference on globally-trained ANN models. On the other hand, spike-based local learning algorithms (e.g. spike-timing dependent plasticity, STDP) have been implemented in spiking neural network (SNN) models \cite{Eshraghian_SNNs_2023} and neuromorphic hardware \cite{Brainscales2010, Loihi2018}.

Here, we highlight how physical learning can be achieved with SOMNs by exploiting their local and global brain-like nonlinear dynamics (including memory), without requiring direct addressability or control of their circuit components. Furthermore, as SOMNs embed networks into their physical hardware, complex neural network algorithms are not required for learning (i.e. ``the hardware is the software'' \cite{laydevant_hardware_2024}).
Two key learning paradigms are described below {and illustrated in Fig.~\ref{fig:Phys-learning}}: physical reservoir computing and associative learning.
{These learning paradigms} demonstrate how the dynamical states of SOMNs, as complex physical systems, can be exploited for learning either in an external output layer, {in the case of reservoir computing,} or within the network itself, {in the case of associative learning}.

\subsection{Physical reservoir computing}\label{sec:physrescomp}

Reservoir computing is a machine learning paradigm where a reservoir (a recurrent neural network) maps an input signal into a higher dimensional manifold that is then analysed by a single output layer, which is the only part that needs to be trained \cite{lukovsevivcius2009reservoir},  {  making this an efficient learning scheme (the weights inside the recurrent neural network are fixed and not trained, while a nonlinear activation function is applied to the nodes).}
Thus, the reservoir may be thought of as extracting features from the data which are then used to learn a model (e.g. regression, classification) in the output layer; the trained model is then used to perform inference on new features generated by the reservoir when presented with unseen data.

In the physical or \textit{in materia} implementation, a nonlinear physical system is used as the reservoir \cite{tanaka2019recent, nakajima2020physical, Usami_2021,liang2024physical,Stepney_2024}. In this context, SOMNs have been demonstrated to satisfy the key requirements of a physical reservoir \cite{Lilak2021,Sheldon2022,sillin2013theoretical, Kuncic_etal_2020, Fu_etal_2020, zhu2021information, milano2022materia, mallinson2024experimental,Mallinson_2023,mallinson2023reservoir, Heywood2024, Steel2025}, endowing \textit{i)} high dimensionality,
with internal dynamics made accessible via multiterminal device configurations; \textit{ii)} high--order nonlinearity, realised through {atomic / }ionic transport mechanisms and recurrent network structure; \textit{iii)} fading memory property (ensuring reservoir states rely only on recent--past inputs and are independent of distant--past inputs)~\cite{Sheldon2022}, as discussed in Sec.~\ref{sec:plasticity}.
{These properties ensure physical reservoirs can generate rich dynamical features from input signals with sufficient expressivity to fit a model to be learned in the output layer.}

In SOMNs, fading memory arises naturally as a consequence of short-term memory of these systems \cite{diaz2019emergent, milano2022materia, Mallinson_2023a,mallinson2024experimental}.

The internal states of SOMNs and their emergent spatiotemporal dynamics~\cite{Barrows2025} are measured via multiterminal electrical measurements ~\cite{demis_2016_nanoarchitectonic,Mallinson2024, pilati2024emerging}. These measurements are typically performed by measuring output currents or floating voltages, or by constructing a conductance matrix where each element corresponds to the conductance between a pair of electrodes.
Fig.~\ref{fig:input-output} shows an example of the input--output mapping by a SOMN reservoir device. In this case, the inputs are sample images from the MNIST handwritten digit dataset that are converted into temporal voltage pulses (red). The output voltages (purple) are read out from device channels (channels 2 and 15 in this example) which produce expressive features for each digit image, with sufficient inter--digit separation to enable classification; the conductance output (green), obtained from the output current, encodes memory within the system and thus can also be leveraged in sequency memory tasks \cite{zhu2023online}. 

Additionally, it has been shown that the spatiotemporal information processing capabilities in multiterminal devices allow an unconventional implementation of physical reservoir computing by exploiting the same electrodes as both inputs and {readouts}, thereby reducing hardware complexity without limiting computing capabilities \cite{milano2023materia}.
{Furthermore, while the output layer is typically trained in software,}
a fully memristive architecture can be realised by coupling SOMNs acting as a reservoir with memristive crossbar arrays \cite{milano2022materia,Lin2025}.
Physical reservoir computing can also be implemented in a two-terminal device by adopting a time multiplexing scheme~\cite{mallinson2023reservoir, milano2025self}.

To date, physical reservoir computing with SOMNs has been experimentally demonstrated across a wide range of tasks, but often exploiting somewhat different operating regimes and readout modalities.
 Tasks that have demonstrated include waveform regression \cite{sillin2013theoretical}, image classification \cite{milano2022materia,zhu2023online}, audio classification \cite{Lilak2021,milano2022speech,Kotooka2024thermally}, time series prediction \cite{mallinson2023reservoir,mallinson2024experimental,milano2023mackey}, non-linear transformations and memory capacity \cite{Usami_2021}, and sequence memory \cite{zhu2023online}, as well as speaker and Braille classification \cite{Steel2025}.

In addition, these experimental results have been complemented by simulations that have been used to test \textit{in silico} various aspects including scalability, the reservoir architecture \cite{Daniels2021, mallinson2023reservoir,Daniels_2023}, the impact of electrode positioning~\cite{Heywood2022}, and the number of electrodes and the dimensionality of the reservoir\cite{daniels2022reservoir}. A  wide range of computing tasks have been simulated including pattern recognition ~\cite{milano2022materia, milano2023materia, Kuncic_etal_2020}, time series prediction ~\cite{milano2023mackey, milano2022materia, Fu_etal_2020, Zhu_transfer2020}, speech recognition ~\cite{milano2022speech}, regression \cite{Hochstetter2021,zhu2021information,Loeffler2021_multitask} and prediction of multidimensional chaotic time series ~\cite{Xu_chaotic_2025}. Furthermore, two-dimensional tracking and autonomous swarm generation has also been simulated \cite{Heywood2024}.

\begin{figure*}[ht]
    \centering
  \includegraphics[width=0.8\linewidth]{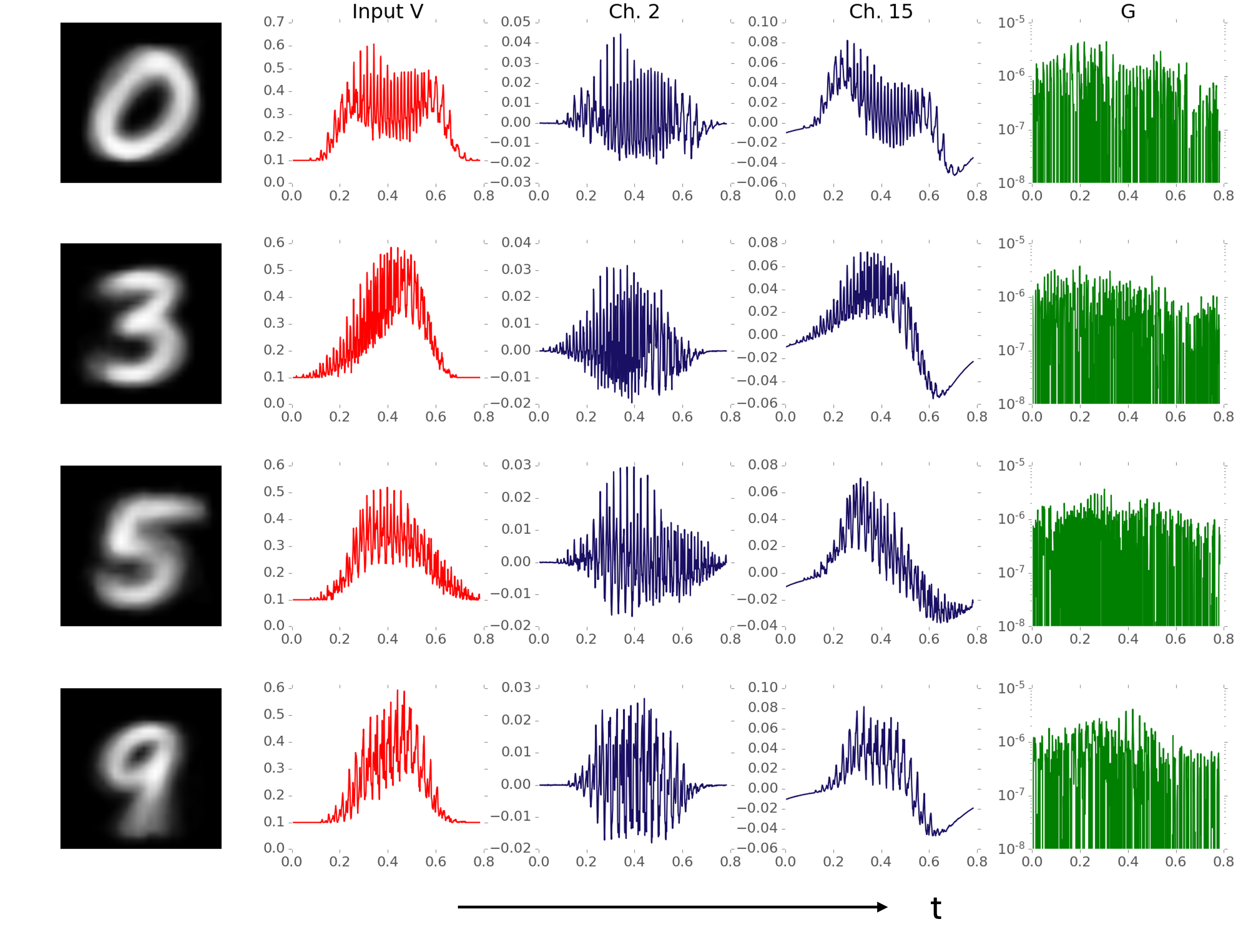}
\caption{Input--output feature mapping for MNIST classification and sequence memory with a nanowire-network physical reservoir. Column~1 shows average images of the digits 0, 3, 5, and 9, each obtained from 100 randomly selected MNIST training samples. Column~2 (red) shows the corresponding temporally encoded input-voltage pulse streams. Columns~3 and~4 (purple) show example voltage readouts from two device channels, which provide digit-dependent dynamical features for the classifier. Column~5 (green) shows the effective conductance $G$, which reflects the evolving internal memory state of the network and is useful for memory-related tasks, rather than classification itself. The figure therefore illustrates how the reservoir transforms input digits into discriminative dynamical features. Figure reproduced from Ref.~\cite{zhu2023online}.}
  \label{fig:input-output}
\end{figure*}

Notwithstanding these attributes, physical reservoir computing {with SOMNs} differs significantly from algorithmic reservoir computing in the following ways: 
(i) SOMNs produce collective memory dynamics by virtue of their memristive elements; and (ii) {SOMN} reservoirs are dynamic networks, with non--static connectivity \cite{Xu_ijcnn2025}; (iii) {SOMN} reservoirs are constrained by Kirchhoff's electrical circuit conservation laws.
{For these reasons, SOMN reservoirs cannot be described mathematically in the same way as standard algorithmic reservoir computers. In conventional reservoir computing, nonlinearity is typically introduced through prescribed activation functions acting on the nodes of a recurrent network with usually random, static connectivity, and performance depends on the optimisation of multiple hyperparameters. By contrast, SOMN reservoirs are governed not only by memristive nonlinearities, but also by Kirchhoff's circuit laws and, more generally, by the physical constraints imposed by the underlying material network and device geometry. Their dynamics therefore emerge from the coupled evolution of internal memory states, transport processes, and constrained network feedback, rather than from an abstract node-based update rule alone. In this sense, SOMNs are not simply hardware implementations of existing reservoir models, but physical systems whose native dynamics can inspire new neural-network models and extensions of the reservoir-computing paradigm \cite{pistolesi2025memristive,Xu_chaotic_2025}.}

Similar to biological systems where functionalities are inherently related to the topology of neuronal networks \cite{suarez2021learning},
higher reservoir computing performance may be achieved using SOMNs with scale-free topology \cite{mallinson2023reservoir}. Higher computational performance can also be achieved experimentally 
 via parallel reservoir architectures, which increase the reservoir dimensionality and richness of its output dynamics \cite{mallinson2024experimental}.
Importantly, it has been shown through simulations that performance in reservoir computing tasks using SOMNs contacted with a restricted number of electrodes is close to the upper bounds achievable when using information from every node \cite{daniels2022reservoir}, a key consideration for the experimental realisation of these systems, where information from each node is not practically accessible.
Such simulations provide additional insight into experimental findings that a large number of electrode readouts may not be necessary to achieve high performance, at least on some tasks~\cite{demis_2016_nanoarchitectonic,zhu2023online,Mallinson_2023a}. Simulations and experimental implementations of more complex reservoir computing tasks are needed to qualify these results further.

Several open questions remain in physical reservoir computing with SOMNs, including the energy consumption for real world tasks, and whether methods can be developed for more self-contained learning that does not rely on training an external layer~\cite{Jaeger_2023}.
Indeed, such an approach could help bridging the gap between physical reservoir computing and associative learning, which is discussed in the next section.

\subsection{Associative learning}\label{sec:physasslearn}

Associative learning, i.e. the formation of associations between stimuli and responses, can be implemented in SOMNs by taking advantage of their adaptive conductance changes (i.e. plasticity, cf. sec.~\ref{sec:plasticity}).
External electrical stimuli applied to SOMNs drive local conductance changes that can lead to global reconfigurations in network states as unique conductance pathways are established for different patterns in the stimuli~\cite{Diaz-Alvarez_associative_2020,Li_etal_2020}. The memory encoded in their conductance states~\cite{milano2023tomography} enables SOMNs to recall, or self--learn, these input--output associations.
In contrast to reservoir computing, where learning occurs in an external layer, associative learning occurs directly inside the SOMN via adjustment of junction conductance states. As repeated stimulation leads to material-level changes in conductance pathways, SOMNs self-organise and recall associations between input patterns and global conductance states.
This is more analogous to biological synaptic plasticity than statistical optimisation.
Teaching SOMNs input--output relationships via feedback, without requiring explicit programming also contrasts sharply with conventional implementations of physical neural networks, which typically rely on direct internal state addressability, for example, to physically implement Hopfield associative memory networks~\cite{Hu_associative_2015}.

As self--adjusting to external stimuli mimics biological learning~\cite{Chialvo1999,Carbajal2022}, associative learning in SOMNs effectively provides a physical and material basis for cognition~\cite{Vahl2024,PershinDiVentra_assoc_2010}.
Experimental and theoretical studies have demonstrated the capacity of SOMNs to emulate core cognitive functions, including associative memory and learning~\cite{PershinDiVentra_assoc_2010,Diaz-Alvarez_associative_2020} as well as working memory~\cite{loeffler2023neuromorphic}.
Figure~\ref{fig:Phys-learning} schematically illustrates the learning loop: input stimuli perturb the SOMN; readouts are compared to targets; and corrective inputs guide reconfiguration through internal plasticity~\cite{Barrows2025unc}.
Moreover, implementation of the $n$--back working memory task has revealed that short--term memory can be consolidated into long--term memory by applying both positive and negative feedback, in a manner similar to contrastive learning~\cite{loeffler2023neuromorphic,Chialvo1999,Movellan_1991_contrastive}. This metaplasticity is enabled by the heterogeneous and recurrent network connectivity~\cite{milano2023tomography}.

Both positive and negative feedback enable SOMNs to iteratively adapt their electrical connectivity and conductance pathways, thereby promoting the emergence of robust functional mappings. This iterative correction process can be viewed, by analogy with genetic algorithms, as a form of evolutionary search in which the agreement between the actual output and the desired target acts as a \textit{fitness signal}: configurations that reduce the output error are effectively reinforced, whereas less successful ones are suppressed or redirected~\cite{Schuman2022evo,Mitchell1996}. In this sense, learning can be interpreted as an adaptive exploration of the network's dynamical state space, driven by input--output feedback, constrained by physical laws, and shaped by environmental cues~\cite{Carbajal2022}.
These mechanisms underscore that learning is not defined by task complexity, but rather by the system dynamics and efficiency of feedback. SOMNs bridge conventional and biological models, offering real--time adaptability, dynamic reconfiguration, and connectivity--driven behaviour~\cite{Chialvo1999, Carbajal2022}. A key open question is how many states can be learned and/or stored in these types of physical architectures?

\section{Outlook}\label{sec:outlook}

The brain demonstrates that intelligence can emerge from the physical properties of matter. This realisation continues to motivate the search for materials and architectures with innate brain-like physical properties that may thus be considered promising substrates for emulating core cognitive functions such as memory and learning. Self-organising memristive networks are compelling candidates in this endeavour. As highlighted in this Perspective, the emergent nonlinear dynamics produced by these non-equilibrium physical systems can be exploited in adaptive learning paradigms, without the need for explicit programming.

This sets the stage for {developing} broader technological applications and {exploring} long-term implications.

Central to their role in physical learning systems is the emergence of global, collective dynamics from relatively simple, local interactions. Nanoscale transport processes (i.e. electron and ionic migration, and atomic rearrangement) continuously modulate junction-level conductance. These local dynamics propagate globally under Kirchhoff constraints and the complex network topology, giving rise to dynamical phase transitions, a hallmark of emergent phenomena.
This interplay between local plasticity and global self-organisation underpins the functional properties that serve to encode information and enable learning.

A key opportunity for future work is to move SOMNs from intriguing functional materials toward \emph{engineerable learning substrates}. We anticipate that this will require physics-grounded modelling pipelines that can begin to connect controllable fabrication parameters and stimulation protocols to task-level performance, although much remains to be understood before such links can be made predictive in practice. Hierarchical and multi-scale approaches, ranging from junction-level models and Kirchhoff-constrained network dynamics to reduced descriptions such as mean-field and stochastic theories, offer a promising route toward this goal. In principle, such approaches could help identify favourable operating regimes---for example in terms of memory depth, nonlinearity, criticality, stability, stimulation strategies, and robustness--energy trade-offs---while also clarifying the collective non-equilibrium physics of disordered adaptive circuits. Establishing these connections is therefore both a fundamental scientific challenge and an exciting practical step toward deploying SOMNs for embedded and continuous learning. In parallel, Equilibrium Propagation, generalized Equilibrium Propagation, and coupled physical learning suggest plausible routes toward in-hardware training, potentially even beyond the quasistatic limit, although substantial theoretical and experimental work is still needed to determine under what conditions such approaches can be implemented reliably and how bias or finite-nudge effects can be controlled~\cite{scellier2017equilibrium,stern2021supervised,laborieux2021scaling,Stern2022,lin2026train}.

Here, we highlight two learning paradigms, reservoir computing and associative learning, that leverage the intrinsic physical dynamics of self-organising memristive networks. Other schemes that are now {under development} include those based on spiking computation\cite{Studholme2023}, probabilistic computing~\cite{Studholme2024, Studholme2025} and energy--based algorithms (e.g. Hopfield networks, Boltzmann machines) that use binary units which may be physically instantiated with thresholded memristive switching. SOMNs that incorporate both spiking neural characteristics and memristive synapses can allow new types of learning behaviour~\cite{Monaghan2025}. The incorporation of molecular synapses into SOMNs is an exciting new avenue for research as the introduction of additional components into the networks (beyond those described in Section \ref{sec:dynamics}) provides new flexibility to tune network properties and opens up myriad new possibilities to implement new styles of computation.

Alongside these emerging computational schemes, an important theoretical frontier is to extend the present modelling and learning framework from predominantly two-terminal descriptions to genuinely \emph{multi-terminal} operation. This extension is especially timely because many of the new functionalities discussed above---including richer spiking behaviour, hybrid synaptic mechanisms, and more flexible forms of computation---are most naturally expressed in architectures that can be driven and read out through several electrodes simultaneously. In realistic experimental SOMNs, multi-terminal configurations enable distributed readout, spatially structured stimulation, and more complex boundary conditions than those captured by an effective two-point conductance description. Developing Kirchhoff-constrained network dynamics, reduced models, and learning rules for such multi-point contacts is therefore a key opportunity for future work, both to broaden the range of accessible tasks and to provide practical design principles for electrode placement, stimulation protocols, and scalability in deployable devices.

A further opportunity is to understand what ultimately sets the upper bound on the information-processing capacity of these memristive networks, including their ability to compress, transform, and encode data. Addressing this question would not only clarify the computational limits of SOMNs, but could also reveal regimes in which the physical substrate itself performs a larger share of the computation. In favourable cases, this may reduce the amount of external training or post-processing required to obtain a desired output, raising the exciting possibility of partially training-free learning schemes that rely more directly on the intrinsic dynamics of the material system~\cite{Smith_2025}.\\
More broadly, SOMNs open a rich design space in which physical structure, adaptive dynamics, and task performance may be co-optimized rather than treated separately. Understanding how network density, topology, material composition, and stimulation protocols shape learning behaviour remains an important challenge, but also a major opportunity for future work. In this respect, SOMNs provide a natural meeting point between neuromorphic engineering, statistical physics, and the complex systems community, where concepts such as emergence, adaptability, and collective computation can be brought to bear on the design of physical learning substrates.\\\ \\
This perspective also connects naturally to the broader idea of guided self-organization~\cite{prokopenko}: rather than prescribing function entirely through explicit programming, one seeks to steer a physical system toward useful dynamical regimes through appropriate constraints, feedback, and environmental coupling. For SOMNs, this suggests an appealing long-term vision in which learning is not simply imposed on the hardware, but arises from the controlled interplay between material properties, network architecture, and stimulation. Developing such principles would deepen our understanding of self-organizing adaptive matter while also providing practical routes toward more autonomous and energy-efficient physical learning systems.
Taken together, these directions suggest that self-organising memristive networks may become more than intriguing neuromorphic analogs. They offer the possibility of a distinct material foundation for intelligent physical systems, in which computation, memory, and adaptation are co-localised in the same substrate. In this future-looking view, SOMNs are not only a platform for implementing brain-inspired ideas, but also an opportunity to discover new forms of computation emerging from disordered, adaptive, and self-organizing matter.

\medskip

\textbf{Acknowledgements.}  F.C.'s work was conducted under the auspices of the National Nuclear Security Administration of the United States Department of Energy at Los Alamos National Laboratory (LANL) under Contract No. DE-AC52-06NA25396, and supported financially via DOE LDRD grant 20240245ER. F.C. is now an employee of Planckian. S.A.B acknowledges support from the MacDiarmid Institute for Advanced Materials and Nanotechnology and the Marsden Fund NZ. G.M and C.R. acknowledge funding by NEURONE, a project funded by the European Union - Next Generation EU, M4C1 CUP I53D23003600006, under program PRIN 2022 (prj code 20229JRTZA). G.M. acknowledges funding by the European Union (ERC, "MEMBRAIN", No. 101160604). Views and opinions expressed are, however, those of the authors only and do not necessarily reflect those of the European Union or the European Research Council. Neither the European Union nor the granting authority can be held responsible for them.

\bibliography{main}

\end{document}